\documentclass[12pt]{article}        %4/8/02

\usepackage{color,graphicx}
\usepackage{latexsym,amssymb,amsmath}
\numberwithin{equation}{section}

\newtheorem{rem}{Remark}[section]

\newcommand{\bsquare}{\hbox{\rule{6pt}{6pt}}}

\newtheorem{theorem}{Theorem}%[section]

\newtheorem{lmm}{Lemma}[section]

\begin{document}

\title{Non-local Markovian symmetric forms on infinite dimensional spaces\\
{\small{Part 2. Examples: non local stochastic quantization of space cut-off quantum fields and infinite particle systems}}}
%\TitleHead{}          %optional
%\dedicatory{and Here is a Dedication}           %optional
\author{
{{Sergio
 \textsc{Albeverio}}}
          \footnote{Inst. Angewandte Mathematik, 
and HCM, Univ. Bonn, Germany, 
email \, :albeverio@iam.uni-bonn.de},
\quad 
{{Toshinao \textsc{Kagawa}}} \footnote{Dept. Information Systems Kanagawa Univ., Yokohama, Japan }, 
\quad
%{\textcolor{blue}{
{{Shyuji \textsc{Kawaski}}} \footnote{Dept. Mathematical Sciences and Phyisics,  Iwate Univ., Morioka, Japan }, 
%}}
\\
 {{Yumi \textsc{Yahagi}}} \footnote{Dept. Mathematical information Tokyo Univ. of Information, Chiba, Japan}, 
\quad
 and \quad
{{Minoru \textsc{W. Yoshida}}} \footnote{Dept. Information Systems Kanagawa Univ., Yokohama, Japan,  
 email:\, washizuminoru@hotmail.com }
 }

\maketitle

\begin{abstract}
The general framework on the  non-local Markovian symmetric forms on 
 weighted $l^p$ $(p \in [1, \infty])$ spaces constructed by [A,Kagawa,Yahagi,Y 2020],  
by restricting  the situation where $p =2$,  is applied to such measure spaces  as the space cut-off $P(\phi)_2$ Euclidean quantum field, 
the $2$-dimensional Euclidean quantum fields with exponential and trigonometric potentials, and the 
field describing a system of an infinite number of classical particles.
 For each measure space, the Markov process corresponding to 
the 
{\it{non-local}} type 
stochastic quantization is constructed.
%Insert your abstract here. Include keywords, PACS and mathematical
%subject classification numbers as needed.
%{\small{\bf{Constructive Euclidean quantum field, Dirichlet form, White noise, 
%Strongly continuoius semi-group, Fock space}}}
%\subclass{MSC code1 \and MSC code2 \and more}
%\end{abstract}
\medskip

\noindent
{\bf{keywords}}:  {\footnotesize{Non local Dirichlet forms on infinite dimensional spaces, space cut-off $P(\phi)_2$,
$\exp \phi$, $\sin \phi$-quantum field models,  Euclidean quantum field,  
infinite particle systems, non-local stochastic quantization.}}
\medskip

\noindent
{\bf{MSC (2020)}}: {\footnotesize{31C25, 46E27, 46N30, 46N50, 47D07, 60H15, 60J46, 60J75, 81S20}}

\end{abstract}

\section{Introduction and preliminaries}

In this paper we apply a general 
 framework on the  non-local Markovian symmetric forms on 
 weighted $l^p$  $(p \in [1, \infty])$ spaces constructed by [A,Kagawa,Yahagi,Y 2020], 
by restricting the situation where $p=2$, i.e., the weighted $l^2$-space,  
to the stochastic quantizations of the space cut-off $P(\phi)_2$ Euclidean quantum field, 
the $2$-dimensional Euclidean quantum fields with exponential and trigonometric potentials, and a field of classical 
(infinite) particle systems. 
Then, for each random field, the Markov process corresponding to 
the 
{\it{non-local}} type 
stochastic quantization is constructed. 
As far as we know, there exists no considerations on the non-local type stochastic quantizations for such random fields  through the arguments by the Dirichlet forms, that are non-local (cf., [A 2003], [A 2016], [A,Ma,R 2015], [A,DeV,Gu1,2], [A,Di Persio, Mastrogiacomo, Smii 2016], 
[A,H-K 76], [A,H-K 77], [A,Kusuoka-sei 2017], [A,R 89], [A,R 90], [A,R 91], [A,R{\"u}diger 2003], [Cat,Chouk 2018], [Da Prato, Debussche 03], [Gu,Ho 2019], [Hairer 2014], 
[Hairer,Mattingly 2016], [Mourrat,Weber 2017], and also for a historical aper{\c c}u on the stochastic quantizations of several random fields cf. [A,Kagawa,Yahagi,Y 2020] and references therein).

In this section, 
we 
first 
recall the abstract results on the non-local Dirichlet forms defined on the Fr{\'e}chet spaces provided in [A,Kagawa,Yahagi,Y 2020], 
and its application to the stochastic quantization of the Euclidean free quantum field which also 
has been considered in [A,Kagawa,Yahagi,Y 2020].
By these preparations, in the next section we proceed to 
construct the solutions of stochastic quantizations corresponding to the space cut-off $P(\phi)_2$ Euclidean quantum field, 
the $2$-dimensional Euclidean fields with exponential and trigonometric potentials, and a field of classical 
(infinite) particle systems.

Here, we limit ourselves to recalling  the results in [A,Kagawa,Yahagi,Y 2020] that will be applied to the stochastic quantizations mentioned above.
Precisely,  for the applications, we restrict our selves to the formulations on 
the weighted $l^2$ spaces and the
 non-local Dirichlet forms with the index 
$0 < \alpha \leq 1$, the index characterizing the order of the 
non-locality which has a corresponcence to the index of the $\alpha$ stable processes.

The {\it abstract} state spaces $S$, on which we define the Markovian symmetric forms, are
the weighted $l^2$ spaces, denoted by  $l^2_{(\beta_i)}$, 
with a given weight $(\beta_i)_{i \in {\mathbb N}}$, $\beta_i \geq 0, 
i \in {\mathbb N}$,  such that
\begin{equation}
S = l^2_{(\beta_i)} \equiv \bigl\{  {\mathbf x} = (x_1, x_2, \dots) \in {\mathbb R}^{\mathbb N} \, : \, 
\| {\mathbf x} \|_{l^2_{(\beta_i)}} \equiv (  \sum_{i=1}^{\infty} {\beta}_i |x_i|^2 )^{\frac12} < \infty    \bigr\}.
\end{equation}

We denote 
by ${\cal B}(S)$ the 
 Borel $\sigma$-field of $S$. 
Suppose that we are given a Borel probability measure $\mu$ on $(S, {\cal {B}}(S))$. 
For each $i \in {\mathbb N}$, 
%{\textcolor{red}{
let  $\sigma_{i^c}$  be the 
sub $\sigma$-field
%}}
 of 
${\cal B}(S)$ that is generated by the Borel sets 
\begin{equation}
B = \left\{ {\mathbf x} \in S \, \, \, \big| \, x_{j_1} \in B_1, \dots x_{j_n} \in B_n \right\}, \quad 
 j_k \ne i, \, \, B_k \in {\cal B}^1, \, \, k=1, \dots, n, \, \, n \in {\mathbb N}, \, \,
\end{equation}
where ${\cal B}^1$ denotes the Borel $\sigma$-field of ${\mathbb R}^1$. Thus, $\sigma_{i^c}$  is the smallest $\sigma$-field that includes every $B$ given by 
(1.2). 
{\it{Namely, ${\sigma}_{i^c}$ is the sub $\sigma$-field of ${\cal B}(S)$ generated by the variables ${\mathbf x} \setminus x_i$, i.e., the variables except of the $i$-th variable $x_i$}}.
For each $i \in {\mathbb N}$, let $\mu(\cdot \, \big| \, \sigma_{i^c})$ be the conditional 
probability,
a one-dimensional probability distribution (i.e., a probability distribution for the $i$-th component $x_i$) valued  $\sigma_{i^c}$ measurable function, 
 that is characterized by (cf. (2.4) of [A,R 91])
\begin{equation}
\mu \big( \{ {\mathbf x} \, \, : \, x_i \in A \} \cap B \big)
= \int_B \mu(A \, \big| \, {\sigma}_{i^c})   \, \mu(d {\mathbf x}),  \quad \forall A \in {\cal B}^1, \, \,
\forall B \in {\sigma}_{i^c}.
\end{equation}
Define
\begin{equation} L^2(S; \mu) \equiv \left\{ f \, \, \,  \Big| \, f : S \to {\mathbb R}, \, {\mbox{measurable and }} \, 
\|f \|_{L^2} = \Bigl( \int_S |f({\mathbf x})|^2 \mu(d {\mathbf x})  \Bigr)^{\frac12} < \infty \right\},
\end{equation}
and
%{\textcolor{blue}{
\begin{equation}
{\cal F}C^{\infty}_0 = \Big\{ f(x_1, \dots, x_n) \cdot \prod_{i \geq 1}I_{\mathbb R}(x_i) \, \, \Big| \, \,  \exists f \in C^{\infty}_0({\mathbb R}^n \to {\mathbb R}), \, n \in {\mathbb N} \, \Big\}
\subset 
L^2(S; \mu),
\end{equation}
%}}
where $C^{\infty}_0({\mathbb R}^n \to {\mathbb R})$ denotes the space of {\it{real valued}} infinitely differentiable functions on ${\mathbb R}^n$ with compact supports. 
%(cf. (4.29) below).  

On $L^2(S;\mu)$, 
for any $0 < \alpha \leq 1$, let us define 
the Markovian symmetric form ${\cal E}_{(\alpha)}$  called  {\it{individually adapted Markovian symmetric form 
{%\textcolor{red}
{of index $\alpha$}} to the measure $\mu$}}, 
the definition of which 
is a natural analogue of  the one for  $\alpha$-stable type 
({\it{non local}})  
Dirichlet form on ${\mathbb R}^d$, $d< \infty$ (cf. (5.3), (1.4)  of  [Fukushima,Uemura 2012], 
also cf.  Remark 1.1 below for the corresponding {\it{local}} Dirichlet forms on $L^2({\mathbb R}^n; \mu)$ with {\it{finite}} $n \in {\mathbb N}$).

For each $0 < \alpha \leq 1$ and  $i \in {\mathbb N}$,
and
for 
the variables 
$$y_i, \, y'_i \in {\mathbb R}^1, \, \, y_i \ne y'_i, \, \,  {\mathbf x}  = (x_1, \dots , x_{i-1},x_i, x_{i+1}, \dots) \in S
\, \, {\mbox{and}}  \, \, 
{\mathbf x} \setminus x_i \equiv (x_1, \dots, x_{i-1},x_{i+1}, \dots),$$
 let 
\begin{eqnarray} 
\lefteqn{
\Phi_{\alpha}(u,v;y_i,y'_i,{\mathbf x} \setminus x_i)
} \nonumber \\
 &&\equiv 
\frac{1}{|y_i- y'_i|^{\alpha +1}} 
\times 
\big\{
u(x_1, \dots, x_{i-1}, y_i, x_{i +1}, \dots) - u(x_1, \dots, x_{i-1}, y'_i, x_{i +1}, \dots) \big\}
 \nonumber \\
&&\times 
\big\{
v(x_1, \dots, x_{i-1}, y_i, x_{i +1}, \dots) - v(x_1, \dots, x_{i-1}, y'_i, x_{i +1}, \dots)
\big\},
\end{eqnarray}
then define
\begin{equation}
{\cal E}_{(\alpha)}^{(i)}(u,v) \equiv 
\int_S
\Big\{
\int_{\mathbb R} I_{\{y\,:\, y \ne x_i\}}(y_i) \,
\Phi_{\alpha}(u,v;y_i,x_i, {\mathbf x}\setminus x_i)  \, {\mu \big( dy_i \, \big| \, \sigma_{i^c} \big) }
\Big\} \mu(d{\mathbf x}),
\end{equation}
%{\textcolor{blue}{
for any $u, v$ such that the right hand side of (1.7) is finite,
where for a set $A$ and a variable $y$, 
$I_A(y)$ denotes the indicator function, and in the sequel, to simplify the notations, we  denote 
$I_{\{y\,:\, y \ne x_i\}}(y_i)$ by, e.g.,  $I_{\{y_i \ne x_i\}}(y_i)$ or $I_{\{ y_i \ne x_i\}}$.

By ${\cal D}_i$, we denote the subset of the space of real valued ${\cal B}(S)$-measurable functions such that 
the right hand side of (1.7) is finite for any $u, \, v \in {\cal D}_i$.
 Let us call $({\cal E}^{(i)}_{(\alpha)}, {\cal D}_i)$ this form, ${\cal D}_i$ being its domain.
Then define 
\begin{equation}
{\cal E}_{(\alpha)}(u,v) \equiv \sum_{i \in {\mathbb N}} {\cal E}_{(\alpha)}^{(i)}(u,v), 
\qquad \forall u, \, v \in \bigcap_{i \in {\mathbb N}} {\cal D}_i.
\end{equation}

%{\textcolor{blue}{
It is easy to see that 
for {\it{the Lipschiz continuous}} functions 
${\tilde{u}} \in C^{\infty}_0({\mathbb R}^n \to {\mathbb R}) \subset {\cal F}C^{\infty}_0$ and 
${\tilde{v}} \in C^{\infty}_0({\mathbb R}^m \to {\mathbb R}) \subset {\cal F}C^{\infty}_0$, $n,m \in {\mathbb N}$, which are representatives of 
$u \in {\cal F}C^{\infty}_0$ and 
$v \in {\cal F}C^{\infty}_0$ respectively,  $n,m \in {\mathbb N}$, 
${\cal E}^{(i)}_{(\alpha)}({\tilde{u}}, {\tilde{v}})$ 
and  ${\cal E}_{(\alpha)}({\tilde{u}}, {\tilde{v}})$ are finite.
Actually,  in [A,Kagawa,Yahagi,Y 2020] it is proved that 
 (1.7) and (1.8) are well defined for ${\cal F}C^{\infty}_0$,
and hence
${\cal F}C^{\infty}_0 \subset \cap_{i \in {\mathbb N}} {\cal D}_i$, 
i.e., it is shown that 
for any real valued ${\cal B}(S)$-measurable function $u$ on $S$,  such that $u=0, \, \mu-a.e.$, it holds that \, 
$\displaystyle{{\cal E}_{(\alpha)}(u,u) =0}$; \, and 
for any $u, v \in {\cal F}C^{\infty}_0$,  there corresponds only one value ${\cal E}_{(\alpha)}(u,v) \in {\mathbb R}$.
Moreover, in [A,Kagawa,Yahagi,Y 2020] it is shown that 
  ${\cal E}_{(\alpha)}$ is a closable Markovian symmetric form.
%}}
Precisely,  the following Theorem 1 holds, which 
 is a restatement of a result given by [A,Kagawa,Yahagi,Y 2020] and  shall  be 
applied to the subsequent discussions in the present paper (in [A,Kagawa,Yahagi,Y 2020], not only  for $0 < \alpha \leq 1$ but also for $0 < \alpha <2$, and for the state spaces  $S$ as weighted $l^p$ spaces, $1 \leq p \leq \infty$, those theorems including the statements corresponding to  Theorems 1, 2 and 3 introduced in this paper are provided):
%}

%{\textcolor{blue}{
\begin{theorem}
{\bf{(The closability)}} \quad
 For the symmetric non-local forms ${\cal E}_{(\alpha)}$, $0 <\alpha \leq 1$ given by (1.8) 
 the following hold:
\\
i) \quad \,  ${\cal E}_{(\alpha)}$ is well-defined on ${\cal F}C^{\infty}_0$; \\
ii) \quad  
$({\cal E}_{(\alpha)},{\cal F}C^{\infty}_0)$ is 
closable in $L^2(S;\mu)$; \\
iii) \quad 
$({\cal E}_{(\alpha)},{\cal F}C^{\infty}_0)$ is 
Markovian.\\
Thus, 
for each $0 < \alpha \leq 1$, the closed extension of $({\cal E}_{(\alpha)},{\cal F}C^{\infty}_0)$ denoted by 
$({\cal E}_{(\alpha)}, {\cal D}({\cal E}_{(\alpha)}))$ with the domain ${\cal D}({\cal E}_{(\alpha)})$,
 is a non-local Dirichlet form on $L^2(S;\mu)$.\\
%\\
%\qquad \qquad \qquad \qquad \qquad \qquad \qquad \qquad \qquad \qquad  \qquad \qquad
%\qquad  \qquad \qquad  \qquad \qquad \quad 
%\bsquare
\end{theorem}

\begin{rem} {} \quad 
In [A,Kagawa,Yahagi,Y 2020], the symmetric form ${\cal E}_{(\alpha)}^{(i)}$ is considered for $0 < \alpha <2$. 
Then, 
the 
{\bf non-local}
symmetric form ${\cal E}_{(\alpha)}^{(i)}$ defined by (1.6) and (1.7), by extending the definition for $0 < \alpha <2$,   can be interpreted as 
{\bf non-local} and 
 {\it \bf local} symmetric forms on 
 the finite dimensional linear space 
$C^{\infty}_0({\mathbb R}^d \to {\mathbb R})$
 (cf., e.g., Example 4 in section 1.2 of [Fukushima 80], and  section II-2 of [M,R 92]), the space of real valued smooth functions with compact supports on 
${\mathbb R}^d$ with some {\it \bf finite} $d \in {\mathbb N}$: \,  
For simplicity, let $d =1$ and  for the Borel probability measure $\mu$, suppose  that there exists a {\it{smooth bounded}} probability density function $\rho \in {\cal S}({\mathbb R} \to {\mathbb R})$,  
an element of Schwartz space of rapidly decreasing functions, 
such that $0 < \rho(x) < \infty$, $\forall x \in {\mathbb R}$.  
Then ${\cal E}_{(\alpha)}^{(i)}$ is interpreted as follows:
$$
\int_{{\mathbb R}^2 \setminus {\mbox{{\footnotesize{"diagonal set"}}}}}
\frac{(f(y) -f(x))(g(y)- g(x))}{|y-x|^{1 + \alpha}} \, \rho(y) \, \rho(x) \, dy \, dx,
\qquad f, \,g \in C_0^{\infty}({\mathbb R} \to {\mathbb R}).
$$
Next, for each $0 < \alpha <2$, let
$$M(\alpha) \equiv (1- \frac12 \alpha)^{\frac{1}{2- \alpha}},$$
then, for $f, \,g \in C_0^{\infty}({\mathbb R} \to {\mathbb R})$, it holds that 
\begin{eqnarray*}
%\lefteqn{
\lim_{\alpha \uparrow 2} \int_{\mathbb R} 
\! \! \! \! &\{& 
\! \! \! \! 
\int_{[x-M(\alpha),\,  x+ M(\alpha)]}
I_{\{y: y \ne x\}}(y') 
\frac{(f(y') -f(x))(g(y')- g(x))}{|y'-x|^{1 + \alpha}} \, \rho(y') \, dy' \} \, \rho(x) \, dx,
\\
%} \\
%&& 
&=& \int_{\mathbb R} f'(x)\, g'(x) \, (\rho (x))^2 \, dx.
\end{eqnarray*}
Also, for each $0 < \alpha <2$ and each $x \in {\mathbb R}$, if we let 
$$M(\alpha; x) \equiv \big( \rho(x)^{-1} (1- \frac12 \alpha) \big)^{\frac{1}{2- \alpha}},$$
then, for $f, \,g \in C_0^{\infty}({\mathbb R} \to {\mathbb R})$, it holds that 
\begin{eqnarray*}
%\lefteqn{
\lim_{\alpha \uparrow 2} \int_{\mathbb R} 
\! \! \! \! &\{& 
\! \! \! \! 
\int_{[x-M(\alpha;x),\,  x+ M(\alpha;x)]}
I_{\{y: y \ne x\}}(y') 
\frac{(f(y') -f(x))(g(y')- g(x))}{|y'-x|^{1 + \alpha}} \, \rho(y') \, dy' \} \,\rho(x) dx,
\\
%} \\
%&& 
&=& \int_{\mathbb R} f'(x)\, g'(x) \, \rho (x) \, dx.
\end{eqnarray*}

Considerations for the {\bf infinite} dimensional situation corresponding to the above {\bf finite} dimensional observation will be carried out in forthcoming work.
\end{rem}

The following theorem is also a part of the main results provided by [A,Kagawa,Yahagi,Y 2020] on the  sufficient conditions (cf. Theorem 2 and Theorem 3 below) under which the Dirichlet forms (i.e. the closed Markovian symmetric forms)  defined above are {\it{strictly quasi-regular}} (cf., [A,R 98, 90, 91] and section IV-3 of [M,R 92], as well as [A 2003] for the meaning of "{\it{strict quasi regular}}" ).

Denote $({\cal E}_{(\alpha)}, {\cal D}({\cal E}_{(\alpha)}))$ the Dirichlet form on $L^2(S;\mu)$, 
with the domain ${\cal D}({\cal E}_{(\alpha)}))$ defined through Theorem 1,   obtained as  the closed extension of 
the closable Markovian symmetric form ${\cal E}_{(\alpha)}$,
 understood as first defined on 
 ${\cal F}C^{\infty}_0$.   We shall use the same notation ${\cal E}_{(\alpha)}$  for the closable form and the closed form.

For each $i \in {\mathbb N}$, we denote by $X_i$ the random variable (i.e., measurable function) on $(S, {\cal B}(S), \mu)$ , that represents the coordinate $x_i$ of  ${\mathbf x} = (x_1, x_2, \dots)$, precisely,
\begin{equation}
X_i \, : S \ni {\mathbf x} \longmapsto x_i \in {\mathbb R}.
\end{equation}
By making use of the random variable $X_i$, we have the following probabilistic expression:
\begin{equation}
\int_S I_{B}(x_i) \, \mu(d {\mathbf x}) = \mu (X_i \in B), \qquad {\mbox{for \, \, $B \in 
{\cal B}(S)$}}.
\end{equation}
\begin{theorem} {\bf{(The strict quasi-regularity)}} \quad
Let  $S= l^2_{(\beta_i)}$, for 
 \,  $0 < \alpha \leq 1$, 
let 
$({\cal E}_{(\alpha)}, {\cal D}({\cal E}_{(\alpha)}))$ be the closed Markovian symmetric form on $L^2(S; \mu)$ given by Theorem 1.
If there exists a 
positive sequence $\{ \gamma_i \}_{i \in {\mathbb N}}$ such that 
$\sum_{i=1}^{\infty} \gamma_i^{-1} < \infty$ (i.e., $\{\gamma^{-\frac{1}2}_i \}_{i \in {\mathbb N}}$ is a positive $l^2$ sequence),
and 
 an  $0 < M_0 <\infty$,
%}}
  and both 
\begin{equation}
\sum_{i=1}^{\infty} 
(\beta_i \gamma_i)^{\frac{1 + \alpha}2}  \cdot
\mu \Big(\beta_i^{\frac{1}2} |X_i| > M_0 \cdot  \gamma_i^{-\frac{1}2}\Big) < \infty,
\end{equation}
\begin{equation}
\mu \Big( \bigcup_{M \in {\mathbb N}} \big\{ |X_i| \leq M \cdot \beta_i^{-\frac{1}{2}}\, \gamma_i^{-\frac{1}2}, \, \forall i \in  {\mathbb N} \big\}  \Big) =1,
\end{equation}
hold, then $({\cal E}_{(\alpha)}, {\cal D}({\cal E}_{(\alpha)}))$ is a  strictly quasi-regular Dirichlet form. 
 \end{theorem}

Next, from [A,Kagawa,Yahagi,Y 2020],  we quote a theorem corresponding to the Markov processes associated to the non-local Dirichlet forms defined above.

Let $({\cal E}_{(\alpha)}, {\cal D}({\cal E}_{(\alpha)}))$, $0 < \alpha \leq 1$,  be the  
family of 
strictly quasi-regular Dirichlet forms 
on $L^2(S; \mu)$ with the state space $S$ 
defined by Theorems 1 and 2.  

For the strictly quasi-regular Dirichlet form 
  $({\cal E}_{(\alpha)}, {\cal D}({\cal E}_{(\alpha)}))$ 
there exists a properly associated
$S$-valued Hunt process (see 
Definitions IV-1.5, 1.8 and 1.13, 
Theorem V-2.13 and Proposition V-2.15 of  [M,R 92]) 
\begin{equation}
{\mathbb M} \equiv \Big(\Omega, {\cal F}, (X_t)_{t \geq 0}, (P_{\mathbf x})_{\mathbf x \in 
S_{\triangle}} \Big).
\end{equation}
$\triangle$ is a point adjoined to $S$ as an isolated point of $S_{\triangle} \equiv S \cup \{\triangle \}$. 
Let $(T_t)_{t \geq 0}$ be the strongly continuous contraction semigroup associated with 
$({\cal E}_{(\alpha)}, {\cal D}({\cal E}_{(\alpha)}))$, and $(p_t)_{t \geq 0}$ be the 
corresponding 
transition semigroup of kernels of the Hunt process $(X_t)_{t \geq 0}$, then for any $u \in {\cal F}C^{\infty}_0 
\subset {\cal D}({\cal E}_{(\alpha)})$ the following holds:
\begin{equation}
\frac{d}{dt} \int_{S} \big(p_t u \big)({\mathbf x}) \, \mu(d {\mathbf x}) 
= \frac{d}{dt} \big( T_t u, 1)_{L^2(S;\mu)} = {\cal E}_{(\alpha)}(T_t u, 1) = 0.
\end{equation}
By this, we see that
\begin{equation}
\int_S \big(p_t u\big)({\mathbf x}) \, \mu(d {\mathbf x}) = \int_S u({\mathbf x}) \, \mu(d {\mathbf x}), 
\quad \forall t \geq 0, \quad \forall u \in {\cal F}C^{\infty}_0,
\end {equation}
and hence, 
by the density of ${\cal F}C^{\infty}_0$ in $L^2(S;\mu)$
\begin{equation}
\int_S P_{\mathbf x} (X_t \in B) \, \mu(d {\mathbf x}) = \mu(B), \qquad \forall B \in {\cal B}(S), \quad \forall t \geq 0.
\end{equation}
Thus, the following  Theorem 3 holds.
\begin{theorem}
{\bf{(Associated Markov process)}} \quad 
Let $0  < \alpha \leq 1$, and let 
$({\cal E}_{(\alpha)}, {\cal D}({\cal E}_{(\alpha)}))$ be a strictly quasi-regular Dirichlet form on 
$L^2(S; \mu)$ that is defined through Theorem 2.
Then to $({\cal E}_{(\alpha)}, {\cal D}({\cal E}_{(\alpha)}))$, there exists a properly associated 
$S$-valued Hunt process ${\mathbb M}$ defined by (1.13), the invariant measure of which is $\mu$ (cf. (1.16)).
%\\ qquad \qquad \qquad \qquad \qquad \qquad \qquad \qquad \qquad \qquad  \qquad \qquad
\qquad  \qquad \qquad  \qquad \qquad \qquad \qquad  \qquad \quad
\bsquare
\end{theorem}

As the final preparations for the main discussions given in the next section, we recall the 
formulation corresponding to the stochastic quantization of the Euclidean free quantum field, discussed in the section 5 of [A,Kagawa,Yahagi,Y 2020].

Let us recall the Bochner-Minlos theorem stated in a general framework.  Let $E$ be a nuclear space ( cf., e.g., Chapters 47-51 of [Tr{\`e}ves 67]). Suppose in particular that $E$ is a countably Hilbert space,   characterized by a sequence of {\it{real}} Hilbert norms $\|{}\, \,{}\|_n$, $n \in {\mathbb N}\cup \{0\}$ such that 
$\|{}\, \,{}\|_0 < \|{}\, \,{}\|_1 < \cdots <\|{}\, \,{}\|_n < \cdots$.
Let $E_n$ be the completion of $E$ with respect to the norm $\|{}\, \,{}\|_n$, then by definition 
$E = \bigcap_{n \geq 0} E_n$ and $E_0 \supset E_1 \supset \cdots \supset E_n \supset \cdots$. Define 
$$E^{\ast}_n \equiv {\mbox{the dual space of $E_n$, and assume 
the identification  $E^{\ast}_0 = E_0$ }}.$$
Then  we have 
$$
E \subset \cdots \subset E_{n+1} \subset E_n \subset \cdots \subset  E_0 = E^{\ast}_0 \subset 
\cdots \subset E^{\ast}_n \subset E^{\ast}_{n+1} \subset \cdots \subset E^{\ast}.
$$
Since by assumption $E$ is a nuclear space, for any $m \in {\mathbb N}\cup\{0\}$ there exists an $n 
\in {\mathbb N}\cup \{0\}$, $n > m$, such that the (canonical) injection $T^n_m\, : E_n \to E_m$ is a trace class (nuclear class) positive operator. The Bochner-Minlos theorem 
 (cf. [Hida 80])
is given as follows:
\begin{theorem}{\bf{(Bochner-Minlos Theorem)}} \\
Let $C(\varphi)$, $\varphi \in E$, be a complex valued function on $E$ such that\\
i) \quad \,\, $C(\varphi)$ is continuous with respect to the norm $\| \, \cdot \, \|_{m}$ for some $m \in {\mathbb N} \cup \{0\}$; \\
ii) \quad \, ({\bf{positive definiteness}}) \quad 
for any $k \in {\mathbb N}$,
$$
\sum_{i,j =1}^k {\bar{\alpha}}_i \alpha_j C({\varphi}_i - {\varphi}_j) \geq 0, \qquad 
\forall \alpha_i \in {\mathbb C}, \, \, \forall {\varphi}_i \in E, \,\,  i=1,\dots, k;
$$
(where ${\bar{\alpha}}$ means complex conjugate of $\alpha$). \\
iii) \quad ({\bf{normalization}}) \quad  $C(0) = 1$. \\
Then, there exists a unique Borel probability measure $\nu$ on $E^{\ast}$ such that 
$$
C(\varphi) = \int_{E^{\ast}} e^{i <\phi, \varphi>} \nu(d \phi), \qquad \varphi \in E.
$$
Moreover,
if the (canonical) injection $T^n_m\, : E_n \to E_m$, for all $ n > m$, is a Hilbert-Schmidt operator, then 
the support of $\nu$ is in $E^{\ast}_n$, where 
$<\phi, \varphi> = {}_{E^{\ast}}<\phi, \varphi>_{E}$ is the dualization between $\phi \in E^{\ast}$ and 
$\varphi \in E$.\\
\qquad \qquad \qquad \qquad \qquad \qquad \qquad \qquad \qquad \qquad  \qquad \qquad
\qquad  \qquad \qquad  \qquad \qquad \quad 
$\Box$
\end{theorem}

\begin{rem} {} \qquad
The assumption on the continuity of $C(\varphi)$ on $E$  given in i) of the above Theorem 4 can be replaced 
by the continuity of $C(\varphi)$ at the {\it{origin}} in  $E$, which is equivalent to i) under 
the assumption that $C(\varphi)$ satisfies ii) and iii) in Theorem 4 (cf. e.g., [It{\^o} K. 76]).
Namely, under the assumption of ii) and iii), the following is equivalent to i): \,
For any $\epsilon >0$ there exists a $\delta >0$ such that 
$$ | C(\varphi) -1 | < \epsilon, \qquad \forall \varphi \in E \quad {\mbox{with}} \quad 
\|\varphi\|_m < \delta.
$$
This can be seen as follows:
Assume that ii) and iii) hold.  For ii), let $k=3$, \,$\alpha_1 = \alpha$, \,$\alpha_2 = -\alpha$, \,$\alpha_3 = \beta$, \,$\varphi_1 = 0$, \,$\varphi_2 = \varphi$ and $\varphi_3 = \psi$, then by the assumption ii),  the 
positive definiteness of $C$, we have
\begin{eqnarray*}
  \lefteqn{
    {\alpha}{\overline{\alpha}} \cdot ( 2C(0) - C(\varphi) - C(- \varphi))
} \\
 &&+ \alpha {\overline{\beta}} \cdot  (C(- \psi - \varphi) - C(- \psi)) 
 + 
    {\overline{\alpha}} \beta \cdot (C(\psi + \varphi) - C( \psi)) + 
\beta {\overline{\beta}} \cdot C(0) \geq 0.
\end{eqnarray*}
By making use of the fact that $C(- \varphi) = {\overline{C(\varphi)}}$, which follows from ii), and the assumption iii), from the above inequality we have
$$
0 \leq \det \left(
      \begin{array}{cc}
          2- C(\varphi) - {\overline{C(\varphi)}} \, \, &{} \, \, {\overline{C(\psi + \varphi) - C(\psi)}}\\
          C(\psi + \varphi) - C(\psi) \, \,  &{} \, \,1
       \end{array}
     \right).
$$
From this it follows that
$$ | C(\psi + \varphi) - C(\psi)|^2 \leq 2 \, |C(\varphi) - 1|.
$$
\qquad \qquad \qquad \qquad \qquad \qquad \qquad \qquad \qquad \qquad  \qquad \qquad
\qquad  \qquad \qquad  \qquad \qquad  \qquad \qquad  
$\bsquare$
\end{rem}

By making use of the support property of $\nu$ by means of the Hilbert-Schmidt 
operators given by Theorem 4, we can present  a framework by which Theorems 1, 2, 3 and 4 
can be applied to the {\it{stochastic quantization}} of  Euclidean quantum fields.

Now, we define an adequate countably Hilbert nuclear space ${\cal H}_0 \supset {\cal S}({\mathbb R}^d \to {\mathbb R}) \equiv {\cal S}({\mathbb R}^d)$, for a given $d \in {\mathbb N}$. \, Let
\begin{equation}
{\cal H}_0 \equiv \Big\{ f \, : \, \|f\|_{{\cal H}_0} = \big((f,f)_{{\cal H}_0} \big)^{\frac12} < \infty, \, \, f : {\mathbb R}^d \to {\mathbb R}, \,\, {\mbox{measurable}} \Big\} \supset  
{\cal S}({\mathbb R}^d),
\end{equation}
where
\begin{equation}
(f,g)_{{\cal H}_0} \equiv (f,g)_{L^2({\mathbb R}^d)} = \int _{{\mathbb R}^d} f(x) g(x) \, dx.
\end{equation}
Let us consider 
the following {\it{pseudo differential operators}} 
on 
${\cal S}({\mathbb R}^d \to {\mathbb R}) \equiv {\cal S}({\mathbb R}^d)$
\begin{equation}
H \equiv (|x|^2 + 1)^{\frac{d+1}2} (- \Delta +1)^{\frac{d +1}2} (|x|^2 + 1 )^{\frac{d +1}2},
\end{equation}
\begin{equation}
H^{-1} \equiv (|x|^2 + 1)^{- \frac{d+1}2} (- \Delta +1)^{- \frac{d +1}2} (|x|^2 + 1 )^{- 
\frac{d +1}2},
\end{equation}
with 
$\Delta$
the $d$-dimensional Laplace operator.
For each $n \in {\mathbb N}$, define 
\begin{equation}
{\cal H}_n \equiv {\mbox{the completion of ${\cal S}({\mathbb R}^d)$ with respect to 
the norm $\|f \|_n$, \, $f \in {\cal S}({\mathbb R}^d)$}},
\end{equation}
where
$\|f \|^2_n = (f,f)_n$ ( in the case where $n=1$, to denote the ${\cal H}_1$ norm we use the exact notation $\| {}\, \, \|_{{\cal H}_1}$,  in order to avoid a confusion between  the notation of some $L^1$ or $l^1$ norms) with 
the corresponding scalar product 
\begin{equation}
(f,g)_n = (H^n f, H^n g) _{{\cal H}_0}, \qquad f, \, g  \in {\cal S}({\mathbb R}^d).
\end{equation}
Moreover we define, for $n \in {\mathbb N}$:
\begin{equation}
{\cal H}_{-n} \equiv {\mbox{the completion of ${\cal S}({\mathbb R}^d)$ with respect to 
the norm $\|f \|_{-n}$, \, $f \in {\cal S}({\mathbb R}^d)$}},
\end{equation}
where
$\|f \|^2_{-n} = (f,f)_{-n}$, with 
\begin{equation}
(f,g)_{-n} = ((H^{-1})^n f, (H^{-1})^n g) _{{\cal H}_0}, \qquad f, \, g  \in {\cal S}({\mathbb R}^d).
\end{equation}
Then  obviously, for  $f  \in {\cal S}({\mathbb R}^d)$, 
\begin{equation}
\|f \|_n \leq \|f \|_{n + 1}, \qquad \|f\|_{-n-1} \leq \|f\|_{-n},
\end{equation}
and by taking the inductive limit and setting ${\cal H} = \bigcap_{n \in {\mathbb N}} {\cal H}_n$, 
we have the following inclusions:
\begin{equation}
{\cal H} \subset \cdots \subset {\cal H}_{n + 1} \subset {\cal H}_n \subset \cdots \subset  {\cal H}_0 \subset  \cdots \subset {\cal H}_{-n} \subset {\cal H}_{-n-1} \subset \cdots \subset {\cal H}^{\ast}.
\end{equation}
The (topological)  dual space of ${\cal H}_n$ is ${\cal H}_{-n}$, $n \in {\mathbb N}$.\\
By the operator 
$H^{-1}$ given  by (1.20) on ${\cal S}({\mathbb R}^d)$ we can define,
on each ${\cal H}_n$, $n \in {\mathbb N}$, the bounded symmetric 
(hence self-adjoint) operators  
\begin{equation}
 (H^{-1})^k, \quad k \in {\mathbb N}\cup\{0\}
\end{equation}
(we use the same notations for the operators on ${\cal S}({\mathbb R}^d)$ and on ${\cal H}_n$).
Hence, for the canonical injection 
\begin{equation}
T^{n +k}_n \, : \, {\cal H}_{n +k} \longrightarrow {\cal H}_n, \qquad k, \,n \in {\mathbb N}\cup \{0\},
\end{equation}
it holds that
$$\| T^{n +k}_{n} f \|_{n} = \| (H^{-1})^k f \|_{{\cal H}_0}, \qquad \forall f \in {{\cal H}_{n +k}},$$
where 
by a simple calculation by means of the Fourier transform, and by Young's inequality, we see that for each $n \in {\mathbb N}\cup \{0\}$, $H^{-1}$ on ${\cal H}_n$ is a Hilbert-Schmidt operator and hence 
$({H}^{-1})^2$ on ${\cal H}_n$ is a trace class operator.

Now,  by applying to the strictly positive self-adjoint 
Hilbert-Schmidt (hence compact) 
operator $H^{-1}$, 
 on ${\cal H}_0 = L^2({\mathbb R}^d \to {\mathbb R})$
  the {\it{Hilbert-Schmidt theorem}} (cf., e.g., Theorem VI 16, Theorem VI 22 of [Reed,Simon 80]) we have that there exists an 
 orthonormal base (O.N.B.) $\{{\varphi}_i\}_{i \in {\mathbb N}}$ of ${\cal H}_0$ such that 
\begin{equation}
H^{-1} {\varphi}_i = \lambda_i \, {\varphi}_i, \qquad i \in {\mathbb N},
\end{equation}
where $\{\lambda_i\}_{i \in {\mathbb N}}$ are the corresponding eigenvalues such that  
\begin{equation}
0 < \cdots < \lambda_2 < \lambda_1 \leq 1,
\quad 
{\mbox{which satisfy}} \quad  \,
\sum_{i \in {\mathbb N}} (\lambda_i)^2 < \infty, \quad {\mbox{i.e.,}} \quad 
\{\lambda_i\}_{i \in {\mathbb N}} \in l^2,
\end{equation}
and $\{{\varphi}_i\}_{i \in {\mathbb N}}$ is indexed adequately corresponding to the finite multiplicity of each $\lambda_i$, $i \in {\mathbb N}$.
By the definition (1.21), (1.22), (1.23) and (1.24) (cf. also (1.27)), for each $n \in {\mathbb N} \cup \{0\}$, 
\begin{equation}
\{(\lambda_i)^{n} {\varphi}_i \}_{i \in {\mathbb N}} \quad {\mbox{is an O.N.B. of ${\cal H}_n$ }}
\end{equation}
and 
\begin{equation}
\{(\lambda_i)^{-n} {\varphi}_i \}_{i \in {\mathbb N}} \quad {\mbox{is an O.N.B. of ${\cal H}_{-n}$ }}
\end{equation}
Thus, 
by denoting ${\mathbb Z}$ the set of integers, 
by the Fourier series expansion 
of functions in ${\cal H}_m$, $m \in {\mathbb Z}$ (cf. (1.21)-(1.24)),
such that 
for $f \in {\cal H}_m$, we have 
\begin{equation}
f  = \sum_{i \in {\mathbb N}} a_i (\lambda_i^{m} {\varphi}_i), \quad {\mbox{with}} \quad 
a_i \equiv \big(f, \, (\lambda_i^{m} {\varphi}_i) \big)_{m}, 
\, \, i \in {\mathbb N}.
\end{equation}
%{\textcolor{blue}{
In particular for $f \in {\cal S}({\mathbb R}^d) \subset {\cal H}_m$, it holds that 
\, $a_i =  {\lambda_i^{-m}} (f, \,
\varphi_i)_{{\cal H}_0}$.
%}}
Moreover we have 
$$
\sum_{i \in {\mathbb N}} a_i^2 = \|f\|_{m}^2,
$$
that yields an {\it{isometric isomorphism}}  $\tau_m$ for each $m \in {\mathbb Z}$ such that 
\begin{equation}
\tau_m \, : \, {\cal H}_m \ni f \longmapsto ({\lambda}_1^m a_1, {\lambda}_2^m a_2, \dots) \in l^2_{(\lambda_i^{-2m})},
\end{equation}
where  $l^2_{(\lambda_i^{-2m})}$ is the weighted $l^2$ space defined by (1.1) with $p=2$, 
and  $\beta_i  = \lambda_i^{-2m}$.
Precisely, for 
$f  = \sum_{i \in {\mathbb N}} a_i (\lambda_i^{m} {\varphi}_i) \in {\cal H}_m$ and 
$g = \sum_{i \in {\mathbb N}} b_i (\lambda_i^{m} {\varphi}_i)  \in {\cal H}_m$, with 
$a_i \equiv \big(f, \, (\lambda_i^{m} {\varphi}_i) \big)_{m}$, 
$b_i \equiv \big(g, \, (\lambda_i^{m} {\varphi}_i) \big)_{m}$, $ i \in {\mathbb N}$, 
by $\tau_m$ the following holds (cf. (1.31) and (1.32)):
$$
(f, \,g)_{m} = \sum_{i \in {\mathbb N}} a_i \cdot b_i = 
\sum_{i \in {\mathbb N} } {\lambda_i^{-m}} (\lambda_i^m a_i) \cdot {\lambda_i^{-m}} (\lambda_i^m b_i) 
 = \big( {\tau}_m f, \, {\tau}_m g \big)_{l^2_{(\lambda_i^{-2m})}}.
$$
By the map $\tau_m$ we can identify, in particular, the two systems of Hilbert spaces 
given by (1.35) and (1.36) through the following diagram:
\begin{equation}
{} \dots \quad
{\cal H}_2 \, \, \subset \, \,  \, \, {\cal H}_1 \, \, \subset {\cal H}_0 = L^2({\mathbb R}^d) \subset {\cal H}_{-1} \subset {\cal H}_{-2} \quad \dots
\end{equation}
{}\qquad \qquad \qquad \qquad \qquad \quad   $\parallel$ \qquad \quad $\parallel$
\qquad \qquad  $\parallel$ \qquad \quad \, \,  $\parallel$ \qquad \, \,  $\parallel$
\begin{equation}
{} \dots \quad
 l^2_{(\lambda_i^{-4})} \, \, \subset  l^2_{(\lambda_i^{-2})} \, \, \subset \, \, \, \, \, \, \, \, l^2 \, \, \, \, \, \, \, \, \,
\subset  \, \, \, \, \, \, \,  l^2_{(\lambda_i^2)} \subset  \, \, \, \, \, \, \,  l^2_{(\lambda_i^{4})} \quad \dots.
\end{equation}
{\bf{Example 0. (The Euclidean free quantum field)}} \quad 
This fundamental example,
which has been considered in  [A,Kagawa,Yahagi,Y 2020]),
 shows how the abstract Theorems 1, 2 and 3, from which we can construct weighted $l^2$-space valued {\it{non-local}} symmetric Markov processes through the {\it{non-local}} Dirichlet forms, can be applied to construct {\it{separable Hilbert space}} 
 (cf. (1.35) and (1.36)) 
valued  
 Markov processes, which is a {\it{stochastic quantization}} of a given {\it{physical}} random field.

Let ${\nu}_0$ be the Euclidean free field 
probability 
measure on ${\cal S}' \equiv {\cal S}'({\mathbb R}^d)$.
It is characterized by 
the (generalized) characteristic function  $C(\varphi)$  in Theorem 4 of $\nu_0$  given by  
\begin{equation}
C(\varphi) = \exp ({- \frac12 (\varphi, (- \Delta + m^2_0)^{-1} \varphi)_{L^2({\mathbb R}^d)}}).
\qquad {\mbox{for \, $\varphi \in {\cal S}({\mathbb R}^d \to {\mathbb R})$}},
\end{equation}
Equivalently, $\nu_0$ is the centered Gaussian probability measure on ${\cal S}'$, the covariance of which 
  is given by
\begin{equation}
\int_{{\cal S}'} <\phi, {\varphi}_1> \cdot  <\phi, {\varphi}_1> \, \nu_0(d \phi) 
= \big(\varphi_1, (- \Delta + m^2_0)^{-1} \varphi_2 \big)_{L^2({\mathbb R}^d)}, 
\qquad \varphi_1, \, \varphi_2 \in {\cal S}({\mathbb R}^d \to {\mathbb R}),
\end{equation}
where $\Delta$ is the $d$-dimensional Laplace operator and $m_0 >0$ ( for $d \geq 3$, we can 
 also 
allow for   $m_0 =0$) is a given mass for this  scalar field.
$\phi (f) = <\phi, f>$, $f \in {\cal S}({\mathbb R}^d \to {\mathbb R})$ is 
 the coordinate process $\phi$ to $\nu_0$ (for the Euclidean free field cf. 
%\textcolor{blue}{
[Pitt 71], 
%}
[Nelson 74], 
 and, e.g., 
[Simon 74], [Glimm,Jaffe 87], 
[A,Y 2002], [A,Ferrario,Y 2004]). By (1.37),  the functional $C(\varphi)$ is continuous with respect to the  norm of the space  ${\cal H}_0 =L^2({\mathbb R}^d)$, and the kernel of 
$(- \Delta + m^2_0)^{-1}$, which is the Fourier inverse transform of $(|\xi|^2 + m^2_0)^{-1}$, $\xi \in {\mathbb R}^d$, is explicitly given by Bessel functions (cf., e.g., section 2-5 of  [Mizohata 73]).
Then, by 
Theorem 4 
and (1.28) 
the support of $\nu_0$ 
can be taken to be 
 in the separable Hilbert  spaces  ${\cal H}_{-n}$, $ n \geq 1$ (cf. (1.35) and (1.36)).

Let us apply Theorems 1, 2 and 3 with $p = \frac12$ to this random field. 
 We start the  consideration from the case where $\alpha= 1$, a simplest situation. 
 Then, we shall state the corresponding results for the cases where $0 < \alpha <1$.

 Now, we take ${\nu}_0$ as a Borel probability measure on ${\cal H}_{-2}$. By (1.34), (1.35) and (1.36), 
by taking $m = -2$, 
$\tau_{-2}$ defines an isometric isomorphism such that 
\begin{equation}
\tau_{-2} \, : \, {\cal H}_{-2} \ni f \longmapsto (a_1, a_2, \dots) \in l^2_{(\lambda_i^{4})},
\quad {\mbox{with}} \quad 
a_i \equiv (f, \, \lambda_i^{-2} {\varphi}_i)_{-2}, \, \, \, i \in {\mathbb N}.
\end{equation}
Define a probability measure $\mu$ on 
$ l^2_{(\lambda_i^{4})}$ such that 
$$\mu(B) \equiv \nu_0 \circ \tau^{-1}_{-2}(B) \quad {\mbox{for}} \quad B \in {\cal B}(l^2_{(\lambda_i^{4})}).
$$
We set $S = l^2_{(\lambda_i^{4})}$ in Theorems 1, 2 and 3, then it follows that the weight $\beta_i$ satisfies $\beta_i = \lambda_i^4$. We can take ${\gamma_i}^{-\frac12} = \lambda_i$ in Theorem 2,  then, from (1.30) we have
\begin{equation}
\sum_{i=1}^{\infty} 
\beta_i \gamma_i \cdot
\mu \Big( \beta_i^{\frac12} |X_i| > M \cdot \gamma_i^{-\frac{1}2}\Big) 
\leq \sum_{i =1}^{\infty} \beta_i \gamma_i = \sum _{i =1}^{\infty} (\lambda_i)^2 < \infty
\end{equation}
(1.40) shows that the condition (1.11) holds.

%{\textcolor{blue}{
Also, as has been mentioned above, 
since $\nu_0({\cal H}_{-n}) = 1$,  for any   $ n \geq 1$, we have 
$$1 = \nu_0({\cal H}_{-1}) = \mu(l^2_{(\lambda_i^2)}) = 
\mu \big( \bigcup_{M \in {\mathbb N}} \{ |X_i| \leq M \beta_i^{- \frac12} \gamma_i^{- \frac12}, \, \forall i \in {\mathbb N} \} \big),
\quad {\mbox{for $\beta_i = \lambda_i^4$, \, $\gamma_i^{- \frac12} = \lambda_i$}}.
$$
This shows that the condition (1.12) is satisfied.
%}}

Thus, by Theorem 2 and Theorem 3,  
for   $\alpha = 1$,
there exists an  $l^2_{((\lambda_i)^{4})}$-valued  Hunt process 
\begin{equation}
{\mathbb M} \equiv \big(\Omega, {\cal F}, (X_t)_{t \geq 0}, (P_{\mathbf x})_{\mathbf x \in 
S_{\triangle}} \big),
\end{equation}
associated  to the non-local Dirichlet form
$({\cal E}_{(\alpha)}, {\cal D}({\cal E}_{(\alpha)}))$.

We can now define an
${\cal H}_{-2}$-valued 
 process 
$(Y_t)_{t \geq 0}$ 
such that
\begin{equation} 
(Y_t)_{t \geq 0} \equiv \big({\tau}^{-1}_{-2}(X_t) \big)_{t \geq 0}.
\end{equation}
Equivalently, by (1.39) for $X_t = (X_1(t), X_2(t), \dots) \in l^2_{(\lambda_i^4)}$, $P_{\mathbf x}-a.e.$, by setting $A_i(t)$ such that 
 $A_i(t) \equiv \lambda_i X_i(t)$  (cf. (1.33) and (1.34)),  we see that  
$Y_t$ \, is also given by 
\begin{equation}
Y_t = \sum_{i \in {\mathbb N}} A_i(t) (\lambda_i^{-2} \varphi_i) = \sum_{i \in {\mathbb N}} X_i(t) \varphi_i \in {\cal H}_{-2}, \qquad \forall t \geq 0, \, \, P_{\mathbf x}-a.e., \, \, 
{\mbox{for any $x \in S_{\triangle}$}}.
\end{equation}
By (1.16) and (1.39), $Y_t$ is an ${\cal H}_{-2}$-valued Hunt process that is a {\it{stochastic quantization}} (according to the definition we gave to this term) with respect to the 
non-local Dirichlet form 
$({\tilde{\cal E}}_{(\alpha)}, {\cal D}({\tilde{\cal E}}_{(\alpha)}))$ 
on $L^2({\cal H}_{-2}, \nu_0)$, that is defined through
$({\cal E}_{(\alpha)}, {\cal D}({\cal E}_{(\alpha)}))$, by making use of $\tau_{-2}$. 
This holds for  $\alpha = 1$.

For the cases where $0 < \alpha <1$, we can also apply Theorems 1, 2 and 3, and then have the corresponding result to (1.43).  For this purpose we have only to notice that for  $\alpha \in (0, 1)$ if we  take $\nu_0$  as a Borel probabilty measure on ${\cal H}_{-3}$, and set 
 $S \equiv l^2_{(\lambda_i^6)}$, $\beta_i \equiv \lambda_i^6$, $\gamma_i \equiv \lambda _i$, and define 
\begin{equation}
\tau_{-3} \, : \, {\cal H}_{-3} \ni f \longmapsto ({\lambda}_1^{-3}a_1, {\lambda}_1^{-3}a_2, \dots) \in l^2_{(\lambda_i^{6})},
\end{equation}
\begin{equation*}
{\mbox{with}} \quad 
a_i \equiv (f, \, \lambda_i^{-3} {\varphi}_i)_{-3}, \, \, \, i \in {\mathbb N},
\end{equation*}
(cf. (1.34), (1.35), (1.36) and (1.39)),
then 
$$\sum_{i=1}^{\infty} (\beta_i \gamma_i)^{\frac{\alpha +1}2} = \sum_{i=1}^{\infty} (\lambda_i)^{2(\alpha +1)}
 < \infty.
$$
As a consequence, 
for $\alpha \in (0, 1)$, 
we then see that  by the this setting (1.11) and (1.12) also hold 
(cf. (1.40) together with the formula given below (1.40)).

Define a probability measure $\mu$ on 
$ l^2_{(\lambda_i^{6})}$ such that 
\begin{equation}
\mu(B) \equiv \nu \circ \tau^{-1}_{-3}(B) \quad {\mbox{for}} \quad B \in {\cal B}(l^2_{(\lambda_i^{6})}).
\end{equation}
Then we have an analogue of (1.43) as follows:
\, 
By Theorem 2 and Theorem 3,  
for each  $0 < \alpha < 1$,
there exists an  $l^2_{(\lambda_i^{6})}$-valued  Hunt process 
\begin{equation}
{\mathbb M} \equiv \big(\Omega, {\cal F}, (X_t)_{t \geq 0}, (P_{\mathbf x})_{\mathbf x \in 
S_{\triangle}} \big), 
\end{equation}
associated  to the non-local Dirichlet form
$({\cal E}_{(\alpha)}, {\cal D}({\cal E}_{(\alpha)}))$.
 By making use of ${\mathbb M}$ we can define an
${\cal H}_{-3}$-valued 
 process 
$(Y_t)_{t \geq 0}$ 
such that 
$(Y_t)_{t \geq 0} \equiv \big({\tau}^{-1}_{-2}(X_t) \big)_{t \geq 0}$,
explicitly, by (1.44) for $X_t = (X_1(t), X_2(t), \dots) \in l^2_{(\lambda_i^6)}$, $P_{\mathbf x}-a.e. \, x \in S_{\triangle}$, by setting $A_i(t)$ such that 
$X_i(t) = \lambda_i^{-3} A_i(t)$  (cf. (1.33) and (1.34)),  then 
$Y_t$ \, is given by 
\begin{equation}
Y_t = \sum_{i \in {\mathbb N}} A_i(t) (\lambda_i^{-3} \varphi_i) = \sum_{i \in {\mathbb N}} X_i(t) \varphi_i \in {\cal H}_{-3}, \qquad \forall t \geq 0, \, \, P_{\mathbf x}-a.e.,
\, x \in S_{\triangle}.
\end{equation}
By (1.16) and (1.44), $Y_t$  is an ${\cal H}_{-3}$-valued Hunt process that is a {\it{stochastic quantization}} with respect to the 
non-local Dirichlet form 
$({\tilde{\cal E}}_{(\alpha)}, {\cal D}({\tilde{\cal E}}_{(\alpha)}))$ 
on $L^2({\cal H}_{-3}, \nu)$, that is defined through
$({\cal E}_{(\alpha)}, {\cal D}({\cal E}_{(\alpha)}))$ (by making use of $\tau_{-3}$ via (1.45)).

The diffusion case $\alpha =2$ was already discussed in [A,R 89],
[A,R 91] (and references therein).
%\qquad \qquad \qquad \qquad \qquad \qquad \qquad \qquad \qquad \qquad  \qquad \qquad
\qquad  \qquad \qquad  \qquad \qquad \quad \qquad \qquad  \qquad \qquad \quad 
\qquad \quad  \qquad \qquad  \qquad \qquad  \qquad \qquad
\bsquare

%\newpage

\section{Other  applications;  quantum field models with interactions, infinite particle systems}
{}\quad {}

Following analogous arguments to 
 the one used for 
Example 0 (see the previous section)
 in the present section we shall treat the following problems, related and complementary to those of [A,Kagawa,Yahagi,Y 2020]:

\noindent
%{\bf{1.\, Non-local type stochastic quantization of the (truncated) H{\o}egh-Krohn model with $d=2$.}} \\
1. \quad 
{\it{Non-local}} type 
stochastic quantization of the (truncated) H{\o}egh-Krohn exponential model with $d=2$ 
(for the considerations on this random field, cf., e.g.,  [H-K 71], [A,H-K 73], [Simon 74], [Fr{\"o}hlich,Park 77], [A,Y 2002], [A,Liang,Zegarlinski 2006], [Kusuoka-shi 92], [A,Kawa,Mih,R 2020], [A,DeV,Gu1,2], [A,Hida,Po,R,Str 89]).

\noindent
%{\bf{2.\, Non-local type stochastic quantization of the (truncated) $P(\phi)_2$ and the Albeverio H{\o}egh-Krohn trigonometric model with $d=2$.}} \\
2. \quad 
{\it{Non-local}} type 
stochastic quantization of the (space cut-off) $P(\phi)_2$ and the Albeverio H{\o}egh-Krohn trigonometric model with $d=2$ (for the considerations on this random field, cf., e.g.,  [A.H-K 73], [A,H-K 79], [Brydges,Fr{\"o}hlich,Sokal 83], [A,H-K,Zegarlinski 89], [A,Y 2002]).

\noindent
%{\bf{3.\, Non-local type stochastic quantization of a field of classical infinite particle system.}}\\
3. \quad  {\it{Non-local}} type 
stochastic quantization of the random fields of classical infinite particle systems 
(for the considerations on this random field, and its {\it{local}} type stochastic quantizations 
by means of local Dirichlet form arguments, cf., e.g., [Ruelle 70], [Osada 96], [Tanemura] 97, [Y 96], [A,Kondratiev,R{\"o}ckner 98]) .

\bigskip
\noindent
{\bf{Example 1. (The space cut-off H{\o}egh-Krohn exponential model with $d=2$.)}} \quad Let ${\nu}_0$ be the Euclidean free field measure on ${\cal S}' \equiv {\cal S}'({\mathbb R}^2)$, 
discussed in Example 0 with $m_0 >0$ (precisely, see (1.37) and (1.38) with $d=2$).
Here, for simplicity we set the mass term $m_0 = 1$.  
Let $a_0$ be a given real number and $g$ a given {\it{positive}} valued function on ${\mathbb R}^2$ 
such that 
\begin{equation}
|a_0| < \sqrt{4 \pi}, \quad  \quad g \in L^2({\mathbb R}^2 \to {\mathbb R}) \cap 
L^1({\mathbb R}^2 \to {\mathbb R}).
\end{equation}
$a_0$ is called "charge" parameter, $g$ (Euclidean) {\it{space cut-off}}.
Define a measurable function 
$V(\cdot)$ 
on the measure space $({\cal S}', {\cal B}({\cal S}'), \nu_0)$, 
 as follows:
\begin{equation}
V(\phi) \equiv V_{a_0. g}(\phi) \equiv \sum_{n=0}^{\infty} \frac{(a_0)^n}{n !} < g, :{\phi}^n:>,
\end{equation}
(often written as \, $:\exp (a_0 <g, {\phi}>):$\, ), 
where $<\cdot, \cdot>$ denotes the dualization between a test function and a distribution, 
and $:{\phi}^n:$ denotes the $n$-th Wick monomial of $\phi$, the ${\cal S}'({\mathbb R}^2 \to {\mathbb R})$-valued random variable of which probability distribution is 
the free field measure 
$\nu_0$ (cf. e.g., [Simon 74]).
Then, it is shown (cf. e.g., [A,H-K 74], [Simon74], [A,Y 2002], the first work in this direction being for the "time zero" version [H-K 74])  
that 
\begin{equation}
V(\phi) \in \bigcap_{p \geq 1} L^p({\cal S}'; \nu_0), \qquad V(\phi) \geq 0, \quad 
\nu_0-a.e..
\end{equation}
By (2.3), since 
\begin{equation}
0 \leq e^{-V(\phi)} \leq 1, \qquad  \nu_0-a.e.,
\end{equation}
it is possible to define a probability measure ${\nu}_{\exp}$ on ${\cal S}'$ such that 
\begin{equation}
{\nu}_{\exp} (d \phi) \equiv 
\frac{1}{Z} e^{- V(\phi)} \, \nu_0 (d \phi),
\end{equation}
where 
$Z$ is the normalizing constant such that 
\begin{equation}
Z \equiv Z_{a_0,g} \equiv 
\int_{{\cal S}'} e^{- V(\phi)} \, \nu_0 (d \phi).
\end{equation}

Now, we shall look at  the support property of the measure ${\nu}_{\exp}$ through the Bochner-Minlos theorem (see Theorem 4 in the previous section), and then apply Theorems 1, 2 and 3 in the previous section quoted from [A,Kagawa,Yahagi,Y 2020], to the random field 
$({\cal S}', {\cal B}({\cal S}'), {\nu}_{\exp})$. To this end, we consider the characteristic function 
$C(\varphi)$ of ${\nu}_{\exp}$:
\begin{equation}
C(\varphi) \equiv 
\int_{{\cal S}'} e^{i <\phi,  \varphi>} \, {\nu}_{\exp} (d \phi), \qquad 
\varphi \in {\cal S}({\mathbb R}^2 \to {\mathbb R}).
\end{equation}
The existence   (i.e., the well definedness) of $C(\varphi)$ 
and its continuity property 
can be guaranteed and shown as follows: \, We first have the following evaluation: 
\begin{eqnarray}
 \lefteqn{ | \int_{{\cal S}'} \sum_{k=0}^{\infty} \frac{(i < \phi, \varphi>)^k}{k !} \,
\nu_{\exp} (d \phi) - 1| }  \nonumber \\
&&= | \frac{1}{Z} \int_{{\cal S}'} \sum_{k=1}^{\infty} \frac{(i < \phi, \varphi>)^k}{k !} \,
e^{- V(\phi)}\, \nu_0(d \phi) |  \nonumber \\
&&\leq
| \frac{1}{Z} \sum_{k =1}^{\infty} \frac{i^k}{k!} \int_{{\cal S}'} < \phi, \varphi>^k \, e^{-V(\phi)} 
\,\nu_0(d \phi) | \\
&&\leq
 \frac{1}{Z} \sum_{k =1}^{\infty} \frac{|i^k|}{k!} \int_{{\cal S}'} |< \phi, \varphi>^k \, e^{-V(\phi)} |
\,\nu_0(d \phi)  \nonumber \\
&&\leq
 \frac{1}{Z} \sum_{k =1}^{\infty} \frac{1}{k!} \int_{{\cal S}'} |< \phi, \varphi>^k |
\,\nu_0(d \phi)  \nonumber \\
&&\leq
\frac{1}{Z} \{ \sum_{l=1}^{\infty} \frac{1}{(2l)!} \, E_{\nu_0} [ <\phi, \varphi>^{2l}] 
+ \sum_{l=1}^{\infty} \frac{1}{(2l-1)!} \, E_{\nu_0}[|<\phi, \varphi>|^l \cdot |<\phi, \varphi>|^{l-1}] \}, \nonumber
\end{eqnarray}
where, 
 (2.5) is used for the first equality, and Fubini Theorem and (2.4) are applied for the first inequality, and 
(2.4) is 
 again 
used for the third inequality, and $E_{\nu_0}[\cdot]$ denotes the expectation with respect to the probability measure $\nu_0$.
Then, by denoting 
$$\|| \varphi \||^2 \equiv ( (- \Delta + 1)^{-1} \varphi, \varphi)_{L^2({\mathbb R}^2)}$$ 
(from the definition of the Euclidean free field, cf., (1.38)) 
for the first term of the right hand side of the last inequality of (2.8), it holds that
\begin{equation}
E_{\nu_0}[ <\phi, \varphi>^{2l} ] = (2l-1)!! \|| \varphi \||^{2l}, \nonumber
\end{equation}
 and for the second term (with $l \geq 2$)   of the right hand side of the last inequality of (2.8), 
by the Cauchy Schwarz inequality, it holds that 
\begin{eqnarray}
 \lefteqn{
    E_{\nu_0}[|<\phi, \varphi>|^l \cdot |<\phi, \varphi>|^{l-1}] } \nonumber \\
  && \leq 
        (E_{\nu_0}[ <\phi, \varphi>^{2l}] \cdot E_{\nu_0}[<\phi, \varphi>^{2l-2}])^{\frac12}
         \nonumber \\
&&=  ((2l-1)!! \|| \varphi \||^{2l} \cdot (2l-3)!! \|| \varphi \||^{2l-2} )^{\frac12},
\nonumber
\end{eqnarray}
also, since,
$$\frac{(2l-1)!!}{(2l)!} = \frac{2^{-l}}{l!} \qquad {\mbox{and}} \qquad 
 \frac{((2l-1)!! \cdot (2l-3)!!)^{\frac12}}{(2l-1)!} \leq \frac{2^{-l+1}}{(l-1)!}, $$
we then see the the right hand side of (2.8) is dominated by 
\begin{equation}
              \frac{1}{Z}  \{ \sum_{l=1}^{\infty} \frac{2^{-l} }{l!} \|| \varphi \||^{2l} + 
E_{\nu_0}[|<\phi, \varphi>|] + \sum_{l=2}^{\infty} \frac{2^{-l+1}}{(l-1)!} \|| \varphi \||^{2l-1} \}
          \leq
              \frac{1}{Z} ( 2 ( e^{\frac12 \|| \varphi \||^2} -1) + \|| \varphi \||).
\end{equation}
(2.8) with (2.9) shows that $C(\varphi)$ is continuous 
 {\it{at the origin}} 
with respect to the norm 
$$\|| \varphi \||^2 \equiv ( (- \Delta + 1)^{-1} \varphi, \varphi)_{L^2({\mathbb R}^2)} \leq \| \varphi \|^2_{L^2({\mathbb R}^2)}, \qquad \varphi \in {\cal S}({\mathbb R}^2 \to {\mathbb R}),$$
and hence, by Remark 1.2,  $C(\varphi)$ has the same continuity as the characteristic function of the Euclidean free field measure given by (1.37) and (1.38) (though,  by  the arguments of  the present evaluation, in (2.9), the sign of the exponent is positive). 
Then, by 
Theorem 4 (with Remark 1.2) 
and (1.28) 
the support of $\nu_{\exp}$ 
can be taken to be 
 in the  Hilbert  spaces  ${\cal H}_{-n}$, $ n \geq 1$ (cf. (1.35) and (1.36)).

Thus, by repeating the same arguments for the Euclidean free field (cf. (1.39)-(1.43)), 
 and 
by 
applying Theorems 1, 2 and 3 in section 1, we have   analogous results on the non-local stochastic quantization for the space cut-off H{\o}egh-Krohn field $\nu_{\exp}$ with $d=2$ to those for the Euclidean free field with $d=2$.

\bigskip
\noindent
{\bf{Example 2. (The space cut-off $P(\phi)_2$ and the Albeverio H{\o}egh-Krohn trigonometric model with ${\mathbf{d=2}}$.)}} \quad
Let us consider once more  the Euclidean free field measure 
 $\nu_0$
on ${\cal S}' \equiv {\cal S}'({\mathbb R}^2 \to {\mathbb R})$, 
discussed in Example 0 (precisely, see (1.37) and (1.38) with $d=2$).
As in Example 1, for simplicity we set the mass term $m_0 = 1$.

Let $\nu_{P(\Phi)}$, $\nu_{\sin}$ and $\nu_{\cos}$ be the probability measures on ${\cal S}'$ that are defined by 
(cf. (2.5))
\begin{equation}
\nu_{P(\Phi)}(d \phi) \equiv \frac{1}{Z_{P(\Phi)}} e^{- V_{P(\Phi)}(\phi)} \nu_0 (d \phi),
\end{equation}
and
\begin{equation}
\nu_{\sin}(d \phi) \equiv \frac{1}{Z_{\sin}} e^{- V_{\sin}(\phi)} \nu_0 (d \phi),
\qquad 
\nu_{\cos}(d \phi) \equiv \frac{1}{Z_{\cos}} e^{- V_{\cos}(\phi)} \nu_0 (d \phi),
\end{equation}
with
\begin{equation*}
V_{P(\Phi)}(\phi) \equiv \lambda \, <g, \, :\phi^{2n}:>,
\end{equation*}
\begin{equation*}
 {\mbox{for some given \quad $\lambda \geq 0$, \, $n \in {\mathbb N}$, 
}}
\end{equation*}
\begin{equation}
{\mbox{
 $g \in L^2({\mathbb R}^2 \to {\mathbb R}) \cap 
L^1({\mathbb R}^2 \to {\mathbb R})$ \,satisfying $g \geq 0$}},
\end{equation}
and
\begin{equation*}
V_{\sin}(\phi) \equiv \lambda \sum_{k=0}^{\infty} \frac{(-1)^k (a_0)^{2k+1}}{(2k +1)!} <g, \, :{\phi}^{2k+1}:>, 
\quad
V_{\cos}(\phi) \equiv \lambda \sum_{k=0}^{\infty} \frac{(-1)^k (a_0)^{2k}}{(2k )!} <g, \, :{\phi}^{2k}:>, 
\end{equation*}
\begin{equation}
 {\mbox{for some given \quad $\lambda \geq 0$, \, $|a_0| < \sqrt{4 \pi}$,
\quad
$g$ \, as in (2.12),}}
\end{equation}
respectively, 
where 
$Z_{P(\Phi)}$, $Z_{\sin}$ and $Z_{\cos}$  are the corresponding normalizing constants such that 
\begin{equation}
Z_{P(\Phi)} \equiv 
\int_{{\cal S}'} e^{- V_{P(\phi)} }\, \nu_0 (d \phi), 
\quad 
Z_{\sin} \equiv 
\int_{{\cal S}'} e^{- V_{\sin} }\, \nu_0 (d \phi), 
\quad 
Z_{\cos} \equiv 
\int_{{\cal S}'} e^{- V_{cos} }\, \nu_0 (d \phi),
\end{equation}
respectively.  
Under the above settings, it is known that (for the polynomial potential case cf., e.g., 
 [Ne 73a], 
[Simon 74], 
%{\textcolor{blue}{
[Fr{\"o} 74]
%}}
[Glimm,Jaffe 84], and for the trigonometric potential case cf. 
 [A.H-K 73], [Fr{\"o},Sei 76], [Fr{\"o}hlich,Park 77], 
[A,H-K 79], [Glimm,Jaffe 84], [A,Y 2002]), 
one has that  \, 
${\displaystyle{ W \in \bigcap_{r \geq 1} L^r({\cal S}'; \nu_0)}}$ \,  for any
\begin{equation}
%{\mbox{any}} \, \, 
W \, =  \, 
V_{P(\Phi)}, \quad V_{\sin}, \quad V_{\cos}, \quad
e^{- V_{P(\Phi)}},
\quad
e^{- V_{\sin}},
\quad 
e^{- V_{\cos}}.
\end{equation}

The bound (2.4), i.e., $0 \leq e^{-V} \leq 1$, 
which holds for $V_{\exp}$ in Example 1, 
does not hold 
for these $V's$ in the present example, but by 
(2.15), 
similar to the case of $V_{\exp}$,
we can show that the 
characteristic functions (cf. Theorem 4) of these probability measures, 
$\nu_{P(\Phi)}$,  $\nu_{\sin}$ and $\nu_{\cos}$ 
 satisfy the {\it{comparable continuity}} as the characteristic function of the Euclidean free field measure.
Thus by Theorem 4 
and (1.28) 
the support of $\nu_{P(\Phi)}$, $\nu_{\sin}$ and $\nu_{\cos}$ 
can be taken to be 
 in the  Hilbert  spaces  ${\cal H}_{-n}$, $ n \geq 1$ (cf. (1.35) and (1.36)).
Then, for the present random fields,    we can repeat the arguments for the Euclidean free field (cf. (1.39)-(1.41)), 
 getting corresponding  results as the the ones 
 of Example 1. 

The corresponding continuity of the characteristic functions of the probability measures  $\nu_{P(\Phi)}$, $\nu_{\sin}$ and $\nu_{\cos}$ can be seen as follows.

Denote by $F(\phi)$, $\phi \in {\cal S}'({\mathbb R}^2 \to {\mathbb R})$, one of the {\it{positive}} random variables 
$
e^{- V_{P(\Phi)}}/Z_{P(\phi)}$, $e^{- V_{\sin}}/Z_{\sin}$ and $e^{- V_{\cos}}/Z_{\cos}$ on the 
probability space 
$({\cal S}', {\cal B}({\cal S}'), \nu_0)$.  Then by (2.15), by applying the H{\"o}lder's inequality (twice) 
(with $p= \frac43$, $q = 4$), we have (cf. (2.8)), 
for $\varphi \in {\cal S} \equiv {\cal S}({\mathbb R}^2 \to {\mathbb R})$,
\begin{eqnarray}
  \lefteqn{ | \int_{{\cal S}'} \sum_{k=0}^{\infty} \frac{(i < \phi, \varphi>)^k}{k!} F(\phi) \, \nu_0(d \phi) - 1|} \nonumber \\
   &&\leq
   \sum_{k=1}^{\infty} \frac{1}{k!} \int_{{\cal S}'} | < \phi, \varphi>^k| \, F(\phi) \, \nu_0(d \phi)
    \nonumber \\
   &&\leq
    \sum_{k =1}^{\infty} \big\{ \big( \int_{{\cal S}'} |<\phi, \varphi>|^{\frac43 k} \,  \nu_0(d \phi) \big)^{\frac34}\,  \big(\int_{{\cal S}'} (F(\phi))^4 \, \nu_0(d \phi) \big)^{\frac14} \big\}
     \nonumber \\
      &&\leq
     \sum_{k =1}^{\infty} \big\{ \big( \int_{{\cal S}'} |<\phi, \varphi>|^{4 k} \,  \nu_0(d \phi) \big)^{\frac14} \, \big(\int_{{\cal S}'} (F(\phi))^4 \, \nu_0(d \phi) \big)^{\frac14} \big\}.
\end{eqnarray}
   Also, by denoting $r \equiv \||\varphi |\| \equiv ((- \Delta + 1)^{-1} \varphi, \varphi)$, note that the following evaluations hold:
\begin{eqnarray}
 \lefteqn{ 
     \frac{1}{k!} ( \int_{{\cal S}'} (<\phi, \varphi>)^{4k} \, \nu_0(d \phi) )^{\frac14} }
       \nonumber \\
     &&= \frac{1}{k!} ((4k -1)!! \, r^{2k})^{\frac14} \leq \frac{1}{k!} \, r^{\frac{k}2} ( (4k)!!)^{\frac14} \nonumber \\
     &&\leq \frac{1}{k!} r^{\frac{k}2} \big(((2k)!!)^2\big)^{\frac14} = \frac{1}{(k!)^{\frac12}} \, (4 r)^{\frac{k}2},
\end{eqnarray}
and
\begin{equation}
(k!)^{- \frac12} \leq 2^{- \frac{k}2} \frac{1}{(l-1)!}, \qquad {\mbox{for \quad $ k = 2l, \, \, l \in {\mathbb N}$}},
\end{equation}
\begin{equation}
(k!)^{- \frac12} \leq 2^{- \frac{k}2} \frac{1}{(l-2)!}, \qquad {\mbox{for \quad $ k = 2l-1, \, \, l \in {\mathbb N}, \, l \geq 2$}}.
\end{equation}
Then, by applying (2.17), (2.18) and (2.19) to the right hand side of (2.16) we see that
\begin{eqnarray}
  \lefteqn{ | \int_{{\cal S}'} \sum_{k=0}^{\infty} \frac{(i < \phi, \varphi>)^k}{k!} F(\phi) \, \nu_0(d \phi) - 1|} \nonumber \\
   &&\leq
     \sum_{k=1}^{\infty} \frac{1}{(k!)^{\frac12}} (4 r)^{\frac{k}2} \cdot \big( \int_{{\cal S}'} (F(\phi) )^4 \, 
\nu_0(d \phi) \big)^{\frac14}  \nonumber \\
    && \leq 
\big\{ (4 r)^{\frac12} + \sum_{l=1}^{\infty} \frac{1}{(l-1)!} \, 2^{-l} (4r)^l
                              + \sum_{l=2}^{\infty} \frac{1}{(l-2)!} 2^{-l + \frac12} (4r)^{l - \frac12} 
\big\}  \cdot
     \big( \int_{{\cal S}'} (F(\phi))^4 \, \nu_0(d \phi) \big)^{\frac14} 
\nonumber \\
     &&=  2 r^{\frac12} \big\{
           1+ e^{2r} r^{\frac12} \big( 1 + 2^{\frac12} r^{\frac12} \big) \big\}
  \cdot \big( \int_{{\cal S}'} (F(\phi))^4 \, \nu_0(d \phi) \big)^{\frac14}.
\end{eqnarray}
(2.14) and (2.15), (2.16) with (2.20) show that the characteristic functions of $\nu_{P(\phi)}$, $\nu_{\sin}$ and $\nu_{\cos}$ are continuous 
{\it{at the origin}} 
with respect to the norm 
$r = \||\varphi |\| \equiv \big((- \Delta + 1)^{-1} \varphi, \varphi \big)^{\frac12}$ for $\varphi \in {\cal S}'$.
Hence, by Remark 1.2, the 
the 
characteristic functions (cf. Theorem 4) of 
$\nu_{P(\Phi)}$,  $\nu_{\sin}$ and $\nu_{\cos}$ 
 satisfy the {\it{comparable continuity}} as the characteristic function of the Euclidean free field measure.

Thus we can apply Theorems 1, 2 and 3 
to the non-local stochastic quantization
of the random fields $\nu_{P(\Phi)}$, $\nu_{\sin}$ and $\nu_{\cos}$, then for these random fields we also get  the analogous statements as the one for the Euclidean free field with $d=2$ and the 
 one for the 
space cut-off  H{\o}egh-Krohn model in Example 1 (cf. (1.39)-(1.43)).

\bigskip
\noindent
{\bf{Example 3. Non-local stochastic quantization for classical infinite particle systems.)}}\\
 \noindent
 In this example we apply Theorems 1, 2 and 3 to the random fields of classical statistical mechanics considered by [Ruelle 70].

On the local type stochastic quantizations for such random fields, the various considerations 
have been 
 already 
made through the arguments of {\it{local Dirichlet forms}} (for the fundamental formulations cf. [Tanemura 97], [Osada 96], [Y 96], [A,Kondratiev,R{\"o}ckner 98]], and for the corresponding extended considerations cf. [Conache 2018], [Osada 2013] and references therein, also cf. [Lang 77] where 
 a first consideration on the stochastic quantization of such a random field through the 
arguments of an infinite  system of stochastic differential equations is presented). But, as far as we know, there exists no considerations on the {\it{non-local}} type stochastic quantizations for such classical particle systems through the arguments by {\it{non-local Dirichlet forms}}.

We first recall the configuration space for the classical infinite particle systems (for an original formulation, cf. [Ruelle 70], and also cf. [Y 96] for their interpretation as a subspace of Radon measures, which will be used in the subsequent discussions of the present example (the notations, e.g., ${\cal Y}$, adopted here are diffent to the ones in [Ruelle 70] and [Y 96])).
Define
\begin{equation*}
{\cal Y} \equiv \big\{ {\mathbb Y} \, |\, {\mathbb Y} :{\mathbb R}^d \to {\mathbb Z}_{+} \, \, {\mbox{such that \, $\sum_{y \in K} {\mathbb Y}(y) < \infty$ \, for any compact $K \subset 
{\mathbb R}^d$}} \big\},
\end{equation*}
\begin{eqnarray}
\lefteqn{
\sigma[{\cal Y}] \equiv {\mbox{ the $\sigma$-field generated by $\{{\mathbb Y} \, | \, \sum_{y \in B} {\mathbb Y}(y) =m \}$}}, } \nonumber \\
&&{} \qquad
{\mbox{
$B$ running over the bounded Borel set of ${\mathbb R}^d$, \, $m \in {\mathbb Z}_{+}$}},
\end{eqnarray}
where $d \in {\mathbb N}$ is a given dimension, and ${\mathbb Z}_{+}$ is the set of non-negative integers, ${\mathbb Z}_{+} \equiv {\mathbb N}\cup \{0\}$.

On the measurable space $({\cal Y}, \sigma[{\cal Y}])$, suppose that we are given a probability measure $\nu$ that satisfies 
(cf. Corollary 2.8, Prop. 5.2 of [Ruelle 70], and cf. also (2.19) of [Y 96]):
\begin{equation}
\nu\big( \bigcup_{N \in {\mathbb N}} U_N \big) =1,
\end{equation}
and for some given $\gamma >0$ and real $\delta$, 
\begin{equation}
\nu \big( U^c_N \big) \leq \sum_{l=0}^{\infty} \big\{ \exp [- (\gamma N^2 - e^{\delta})] 
\big\}^{l+1},
\end{equation}
where, for $N \in {\mathbb N}$,
\begin{equation}
U_N \equiv \big\{ {\mathbb Y} \in {\cal Y} \, |\, \forall l \in {\mathbb Z}_{+}, \,
\sum_{r: |r| \leq l} n({\mathbb Y}, r)^2 \leq N^2 (2l + 1)^d \big\}
\end{equation}
with
\begin{equation}
n({\mathbb Y}, r) \equiv \sum_{y \in Q_r} {\mathbb Y}(y),
\end{equation}
\begin{equation}
Q_r \equiv \big\{ y = (y^1, \cdots, y^d) \in {\mathbb R}^d \, | \, r^j -\frac12 \leq y^j < r^j + \frac12, \, j=1, \dots, d \big\},
\quad 
r= (r^1, \dots, r^d) \in {\mathbb Z}^d.
\end{equation}

Define (cf. (2.22))
\begin{equation}
{\cal Y}_0 \equiv \bigcup_{N \in {\mathbb N}} U_N \subset {\cal Y},
\end{equation}
and the corresponding $\sigma$-field 
\begin{equation}
\sigma [{\cal Y}_0] \equiv \{ B \cap {\cal Y}_0 \, | \, B \in \sigma [{\cal Y}] \}.
\end{equation}
By (2.22) and (2.23), we restrict $\nu$ (originally defined on ${\cal Y}$) to ${\cal Y}_0 \subset {\cal Y}$, and denote the restriction by the same notation $\nu$.
Then, we can define the probability space 
\begin{equation}
({\cal Y}_0, \sigma [{\cal Y}_0], \nu).
\end{equation}

Subsequently, we shall interpret the probability measure $\nu$ on the configuration space ${\cal Y}_0$ to a probability measure on a subset of the space of  Radon measures on ${\mathbb R}^d$. 
We note that 
 each ${\mathbb Y} \in {\cal Y}$ can be identified with $z$, an element of positive integer valued Radon measures on ${\mathbb R}^d$, such that (cf. [Y 96])
\begin{equation}
z = \sum_{i=1}^{\infty} m_i {\delta}_{y_i}, \quad y_i \in {\mathbb R}^d, \quad i \in {\mathbb N},
\end{equation}
for given ${\mathbb Y} \in {\cal Y}$ with 
$\{y_i\}_{i \in {\mathbb N}} \equiv \{ y \in {\mathbb R}^d \, | \, {\mathbb Y} (y) \ne 0 \}$ and 
$m_i = {\mathbb Y} (y_i)$, where $\delta_{y_i}$ denotes the Dirac measure at the point $y_i \in {\mathbb R}^d$.
Define 
\begin{equation}
{\tilde{\cal Y}}_0 \equiv \big\{ z \, | \, z \, {\mbox{corresponds with an ${\mathbb Y} \in {\cal Y}_0$ by (2.30)}} \big\},
\end{equation}
\begin{equation}
\sigma[{\tilde{\cal Y}}_0] \equiv \Big\{  \big\{ z \, | \, z \, {\mbox{corresponds with an ${\mathbb Y} \in B$,   by (2.30)}} \big\} \, \Big| \, 
B \in \sigma[{\cal Y}_0]  \Big\},
\end{equation}
and 
\begin{equation}
{\tilde{\nu}}({\tilde{B}}) \equiv \nu(B), 
\quad {\mbox{for}} \, \,  {\tilde{B}} = 
 \big\{ z \, | \, z \, {\mbox{corresponds with an ${\mathbb Y} \in B$,   by (2.30)}} \big\}  \in \sigma[{\tilde{\cal Y}}_0].
\end{equation}
Then, from (2.29), through (2.30)-(2.33), 
 we can 
define the probability space (on a {\it{subset}} of the space of Radon measures 
(cf., e.g., [Kallenberg 83],  also as a general reference cf. [Tr{\`e}ves 67] ) such that
\begin{equation}
({\tilde{\cal Y}}_0, \sigma [{\tilde{\cal Y}}_0], {\tilde{\nu}}).
\end{equation}

Next, we embed ${\tilde{\cal Y}}_0$ defined by (2.31) into a Hilbert space, and interpret the present random field $({\tilde{\cal Y}}_0, \sigma [{\tilde{\cal Y}}_0], {\tilde{\nu}})$ to be the one 
on which we can apply Theorems 1, 2 and 3.  To this end, 
for the present consideration, we modify the Hilbert-Schmidt operator and the corresponding nuclear space defined through (1.19)-(1.32)  as follows:
Let 
\begin{equation}
{\cal H}_0 \equiv L^2({\mathbb R}^d \to {\mathbb R}; \lambda), \quad 
{\mbox{with $\lambda$ the Lebesgue measure on ${\mathbb R}^d$}},
\end{equation}
and
\begin{equation}
{\tilde{H}} \equiv (|x|^2 + 1)^{d +1} (- \Delta + 1)^{\frac{d +1}2} (|x|^2 + 1)^{d +1},
\end{equation}
\begin{equation}
{\tilde{H}}^{-1} \equiv (|x|^2 + 1)^{-(d +1)} (- \Delta + 1)^{-\frac{d +1}2} (|x|^2 + 1)^{-(d +1)},
\end{equation}
${\tilde{\cal H}}_n$ and ${\tilde{\cal H}}_{-n}$ be  the completion of ${\cal S} \equiv {\cal S}({\mathbb R}^d \to {\mathbb R})$, the space of  Schwartz's rapidly decreasing functions, equipped with the norms corresponding to the inner products $(\cdot, \cdot)_n$ and $(\cdot, \cdot)_{-n}$ respectively  such that 
\begin{equation}
(f,g)_n \equiv ({\tilde{H}}^n f, {\tilde{H}}^n g)_{{\cal H}_0}, \qquad 
f, \, g \in {\cal S},
\end{equation}
\begin{equation}
(f,g)_{-n} \equiv (({\tilde{H}}^{-1})^n f, ({\tilde{H}}^{-1})^n g)_{{\cal H}_0}, \qquad 
f, \, g \in {\cal S}.
\end{equation}
Then, through  arguments 
 analogous to those 
performed in section 1 and in the previous examples 
(with obvious adequate modifications of notations and notions) we have the {\it{continuous}} inclusion
\begin{equation}
{\tilde{\cal H}}_3 \subset
{\tilde{\cal H}}_2 \subset {\tilde{\cal H}}_1 \subset 
{\cal H}_0 = L^2 \subset {\tilde{\cal H}}_{-1} \subset {\tilde{\cal H}}_{-2}
\subset {\tilde{\cal H}}_{-3},
\end{equation}
and by the self-adjoint extension of $H^{-1}$ on ${\cal S}$, for each domain 
${\tilde{\cal H}}_n$, $n \in {\mathbb Z}$ (setting ${\tilde{\cal H}}_0 = {\cal H}_0$), 
we also have 
the Hilbert-Schmidt operator ${\tilde{H}}^{-1}$.

Let $\{{\tilde{\varphi}}_i\}_{i \in {\mathbb N}}$ be the orthonormal base of the Hilbert space ${\cal H}_0$ (cf. (1.29)-(1.32)) such that 
\begin{equation}
{\tilde{H}}^{-1} {\tilde{\varphi}}_i = {\tilde{\lambda}}_i {\tilde{\varphi}}_i, \qquad i \in {\mathbb N},
\end{equation}
where $\{ {\tilde{\lambda}}_i \}_{i \in {\mathbb N}}$ is the family of the corresponding eigenvalues, that satisfies
\begin{equation}
0 <\cdots <  {\tilde{\lambda}}_2 < {\tilde{\lambda}}_1 \leq 1, \qquad  \qquad  \quad
\{ {\tilde{\lambda}}_i \}_{i \in {\mathbb N}} \in l^2.
\end{equation}

Through the preparations (2.35)-(2.42) above, we see that for any ${\mathbb Y} \in {\cal Y}_0$ the corresponding Radon measure $z\in {\tilde{\cal Y}}_0$ defined by (2.30) satisfies 
\begin{equation}
z \in {\tilde{\cal H}}_{-1}.
\end{equation}
Equivalently, we are able to show that the following Lemma 2.1 holds:
\begin{lmm}
For the subset of Radon measures ${\tilde{\cal Y}}_0$ defined by (2.31), it holds that 
\begin{equation*}
{\tilde{\cal Y}}_0  \subset {\tilde{\cal H}}_{-1}.
\end{equation*}
\end{lmm}

The proof of Lemma 2.1 is given in Section 3 Appendix.

Moreover,  the following Lemma 2.2 holds.
\begin{lmm}
For the $\sigma$-field $\sigma[{\tilde{\cal Y}}_0]$ defined by (2.32) and the Borel field 
${\cal B}({\tilde{\cal H}}_{-r})$ of the Hilbert space ${\tilde{\cal H}}_{-r}$, it holds that
$$\sigma[{\tilde{\cal Y}}_0] \supset \big( {\cal B}({\tilde{\cal H}}_{-r}) \cap {\tilde{\cal Y}}_0 \big), \qquad 
{\mbox{for \,$r \geq 1$}}.
$$
\end{lmm}

The proof of Lemma 2.2 is also given in the Appendix.

By Lemmas 2.1 and  2.2, the {\it{probability measure}} 
of the classical infinite particle system
${\tilde{\nu}}$ on $({\tilde{\cal Y}}_0, \sigma[{\tilde{\cal Y}}_0])$ 
(cf. (2.34)) can be extended, 
for each $r \geq 1$, 
to a {\it{Borel probability measure}} ${\nu}_r$ on $({\tilde{\cal H}}_{-r}, {\cal B}({\tilde{\cal H}}_{-r} )$ 
 as follows: 
\begin{equation}
{\nu}_r(B) = {\tilde{\nu}}(B \cap {\tilde{\cal Y}}_0), \qquad B \in  {\cal B}({\tilde{\cal H}}_{-r}).
\end{equation}

For the subsequent discussion, we take
 $r =3$, and consider the corresponding extended random field 
$({\tilde{\cal H}}_{-3}, {\cal B}( {\tilde{\cal H}}_{-3}), {\nu}_3)$
 to  $({\tilde{\cal Y}}_0, \sigma [{\tilde{\cal Y}}_0], {\tilde{\nu}})$  (cf. (2.34)).

By making use of Lemma 2.2, we shall proceed to the application of Theorems 1, 2 and 3 to the random field $({\tilde{\cal H}}_{-3}, {\cal B}( {\tilde{\cal H}}_{-3}), {\nu}_3)$.  From (2.40), (2.41) and (2.42) (cf. (1.29)-(1.34)), we see that 
for $k = 0, 1, 2, 3$ with ${\tilde{\cal H}}_{-0} = {\tilde{\cal H}}_{0}  = {\cal H}$,   
\begin{equation}
\{({\tilde{\lambda}}_i)^k {\tilde{\varphi}}_i \}_{i \in {\mathbb N}} 
\qquad {\mbox{is an O.N.B. of ${\tilde{\cal H}}_k$}}, 
\end{equation}
\begin{equation}
\{({\tilde{\lambda}}_i)^{-k} {\tilde{\varphi}}_i \}_{i \in {\mathbb N}} 
\qquad {\mbox{is an O.N.B. of ${\tilde{\cal H}}_{-k}$,    }}, 
\end{equation}
and we can define an isometric isomorphism $\tau$ between ${\tilde{\cal H}}_{-3}$ and 
$l^2({\tilde{\lambda}}^6)$ such that 
\begin{equation}
\tau \, : \, {\tilde{\cal H}}_{-3} \ni f \longmapsto ({\tilde{\lambda}}^{-3}_1 a_1, 
{\tilde{\lambda}}^{-3}_2 a_2, \dots) \in l^2({\tilde{\lambda}}^6),
\quad {\mbox{with \, $a_i \equiv (f, {\tilde{\lambda}}^{-3}_i {\tilde{\varphi}}_i )_{-3}, \, i \in {\mathbb N}$, }}
\end{equation}
where the inner product $(\cdot, \cdot)_{-3}$ is defined, not by (1.24), but by (2.39).
Then, through $\nu_p$ defined by Lemma 2.2 and the mapping $\tau$, we can define a Borel probability measure $\mu$ on $l^2({\tilde{\lambda}}^6)$ as follows:
\begin{equation}
\mu(B) \equiv {\nu}_3 \circ {\tau}^{-1} (B) \qquad {\mbox{for \, $B \in {\cal B}(l^2({\tilde{\lambda}}^6))$}}.
\end{equation}
Now, by setting $S = l^2({\tilde{\lambda}}^6)$ in Theorems 1, 2 and 3, for each $\alpha \in (0, 1]$, we have an $l^2({\tilde{\lambda}}^6)$-valued Hunt process 
${\mathbb M} \equiv (\Omega, {\cal F}, (X_t)_{t \geq 0}, (P_{\mathbf x})_{\mathbf x \in 
S_{\triangle}} )$, associated to the non-local Dirichlet form $({\cal E}_{(\alpha)}, {\cal D}({\cal E}_{(\alpha)}))$ on $L^2(S, \mu)$. We can then define an 
${\tilde{\cal H}}_{-3}$-valued Hunt process $(Y_t)_{t \geq 0}$ (cf. (1.42) and (1.43)) such that 
\begin{equation}
Y_t = \left\{
       \begin{array}{ll}
          \sum_{i \in {\mathbb N}}X_i(t) {\tilde{\varphi}}_i, &
            \qquad X_i(t) \ne \Delta \\[0.2cm]
         {\Delta}', & \qquad X_i(t) = \Delta,
       \end{array} \right. 
\end{equation}
where ${\Delta}'$ is a point adjoint to  ${\tilde{\cal H}}_{-3}$ (the cemetery).

\begin{rem} {} \quad i) \,  These considerations performed in Example 3 is adapted to all the case $\alpha \in (0, 1]$, 
but, if we restrict our discussions to the case where $\alpha =1$, then 
we are able to take  
 $S = l^2({\tilde{\lambda}}^4)$, and have an $l^2({\tilde{\lambda}}^4)$-valued Hunt process 
${\mathbb M} \equiv (\Omega, {\cal F}, (X_t)_{t \geq 0}, (P_{\mathbf x})_{\mathbf x \in 
S_{\triangle}} )$, associated to the non-local Dirichlet form $({\cal E}_{(\alpha)}, {\cal D}({\cal E}_{(\alpha)}))$ on $L^2(S, \mu)$, and then  we can  define an 
${\tilde{\cal H}}_{-2}$-valued Hunt process $(Y_t)_{t \geq 0}$ (cf. (2.49)) through the same discussion as Example 0.
\medskip 

\noindent
ii) \, 
In order to consider 
 another 
 jump type Markov processes, which are natural analogues of the diffusion process, with invariant measure  ${\tilde{\nu}}$ defined by (2.33) and (2.34), constructed through the local type Dirichlet form defined by [Y 96], 
where the present ${\tilde{\nu}}$ was denoted by $\mu$,
 we should define, analogous to [Y 96],   a corresponding non-local type Dirichlet forms   by making use of 
a  {\it{system of density distributions}} 
(cf. Assumption 1 and Remark 1 of [Y 96]).

These 
would be different from the non-local type Dirichlet forms discussed in the present paper, since they would 
involve the mentioned system of density distributions.
\end{rem}

\section{Appendix}
{}\quad {}

\noindent
{\bf{Proof of Lemma 2.1.}} \quad 
From (2.22) and (2.23), the assumption for the original measure $\nu$, Lemma 2.1 can be proven as follows: \, It suffices to show that (2.43) holds for any $z$ that corresponds with an ${\mathbb Y} \in U_N \subset {\cal Y}_0$ for some $N \in {\mathbb N}$ (see (2.27)).

\begin{equation}
{\mbox{For \quad 
${\mathbb Y} \in U_N \subset {\cal Y}_0$ \, \, with \, \, $N \in {\mathbb N}$, \qquad let \quad
$z = \sum_{i=1}^{\infty} m_i \delta_{y_i} \in {\tilde{\cal Y}}_0$,}}
\end{equation}
the Radon measure corresponding with ${\mathbb Y}$ through (2.30).
Then,
for any test function $\varphi \in {\cal S}$ \,(denoting by $<z, \varphi>$ the dualization between the distribution $z$ and the test function $\varphi$), we have
\begin{eqnarray}
  \lefteqn{
      |<z,\varphi>| = |\sum_{i =1}^{\infty} m_i \varphi(y_i) | = | \sum_{l =0}^{\infty} \big( \sum_{y_i \in Q_r: |r| =l} m_i \varphi(y_i) \big)| }   \nonumber \\
     &&\leq \sum_{l=0}^{\infty} \big( \sum_{|r| = l} n({\mathbb Y}, r) (\sup_{y \in  Q_r} |{\varphi}(y)|) \big) 
  \leq \sum_{l=0}^{\infty} \big( \sum_{|r| = l} (n^2({\mathbb Y}, r)) (\sup_{y \in  Q_r} |{\varphi}(y)|) \big)  \nonumber \\
      &&\leq \sum_{l=0}^{\infty} \big( \sum_{|r| = l} (n^2({\mathbb Y}, r)) (\sup_{y \in  Q_r} 
| (|y|^2 + 1)^{d + 1} {\varphi}(y)|) \big) \nonumber \\
      &&\leq \sum_{l=0}^{\infty} \Big( \sum_{|r| = l} (n^2({\mathbb Y}, r))
        \big( \int_{Q_r} ((- \Delta + 1)^{\frac{d+1}2} (|y|^2 + 1)^{d +1} \varphi (y) )^2 dy \big)^{\frac12} \Big)   \nonumber \\
      &&\leq 
             \sum_{l=0}^{\infty} \Big( \sum_{|r| = l} (n^2({\mathbb Y}, r)) 
        \big( \int_{Q_r} ((|y|^2 + 1)^{-(d+1)} )^2 dy \big)^{\frac12} \nonumber \\
      &&\times  \big( \int_{Q_r} \big((|y|^2 + 1)^{d+1} ((- \Delta + 1)^{\frac{d+1}2} (|y|^2 + 1)^{d +1} \varphi (y) \big)^2 dy \big)^{\frac12} \Big),  \nonumber \\
    &&\leq \sum_{l=0}^{\infty} \Big( \sum_{|r| = l} (n^2({\mathbb Y}, r)) 
        \big( \int_{Q_r} ((|y|^2 + 1)^{-(d+1)} )^2 dy \big)^{\frac12} \Big)
 \|\varphi \|_{{\tilde{\cal H}}_1}.
\end{eqnarray}
In the above deductions, to get the third inequality we have applied the Sobolev's embedding theorem (cf., e.g., [Mizohata 73])  that gives
for the Sobolev space $W^{m,2}$ with $m=[\frac{d}2] +1$, 
that 
$W^{m,2} \subset C_{b}({\mathbb R}^d)$ (cf. the explanation given below (3.10)),  
where $C_b({\mathbb R}^d)$ denotes 
the space of {\it{real}} valued bounded continuous functions on ${\mathbb R}^d$.  Since, 
for $r \in {\mathbb Z}^d$ 
by denoting  $|r| = l$,  for some $C < \infty$  it holds that 
\begin{equation}
\int_{Q_r}  \big( (|y|^2 + 1)^{-(d +1)} \big)^2 dt 
 \leq C (l^2 +1 )^{- \frac{3d +5}2}, 
\end{equation}
by this together with  (3.1), we can evaluate the right hand side of (3.2).
We consequently see that the following holds for some constants $C_1, \, C_2, C_3 < \infty$ \, 
(only $C_3$ depends on $N$) :
\begin{eqnarray}
  \lefteqn{ |<z,\varphi>| =
      C_1 \big\{ \sum_{l=1}^{\infty} n^2({\mathbb Y}, r) (2l + 1)^{-d} (l^2 +1)^{- \frac{3d +5}4 + \frac{d}2} + N^2 \big\} \|\varphi \|_{{\tilde{\cal H}}_1} } \nonumber \\
  &&\leq 
  C_2 \big\{ \sum_{l=1}^{\infty} N^2 (l+ 1)^{- \frac{d}2 - \frac52} + N^2 \big\} \|\varphi \|_{{\tilde{\cal H}}_1} 
                                \leq C_3 \|\varphi \|_{{\tilde{\cal H}}_1}, \qquad \forall \varphi \in {\tilde{\cal H}}_1.
\end{eqnarray}
(3.4) shows that 
\begin{equation}
z \in{\tilde{\cal H}}^{*}_1 = {\tilde{\cal H}}_{-1}, 
\end{equation}
for any $z$ that corresponds with an ${\mathbb Y} \in U_N \subset {\cal Y}_0$ for some $N \in {\mathbb N}$.  Since $N \in {\mathbb N}$ is arbitrary, 
 the proof of (2.43) is completed. 
\qquad  \qquad \qquad  \qquad \qquad \quad 
 \qquad \qquad  \qquad \qquad \qquad \quad \qquad
\bsquare

\bigskip
\noindent
{\bf{Proof of Lemma 2.2.}} 
\qquad
Let $r \geq 1$. 
We shall show that 
\begin{equation}
\sigma[{\tilde{\cal Y}}_0] \supset \big({\cal B}({\tilde{\cal H}}_{-r}) \cap {\tilde{\cal Y}}_0 \big).
\end{equation}
Since ${\tilde{\cal H}}_{-r}$ is a separable Hilbert space, and hence, it is a {\it{Souslin space}}, and also since the dual space of  ${\tilde{\cal H}}_{-r}$ is ${\tilde{\cal H}}_{r}$ 
(see (2.40)), it holds that (cf. e.g., [Ba 70])
\begin{equation}
{\cal B}({\tilde{\cal H}}_{-r}) = \sigma[{\tilde{\cal H}}_{r}] \equiv 
{\mbox{the $\sigma$-field generated by 
$\{z \,|\, z \in {\tilde{\cal H}}_{-r}, \varphi(z) <t\}, \, t \in {\mathbb R}, \, \varphi \in {\tilde{\cal H}}_{r}$}},
\end{equation}
where $\varphi(z) = <z, \varphi>$ denotes the dualization between the distribution 
$z \in {\tilde{\cal H}}_{-r}$ and the test function $\varphi \in {\tilde{\cal H}}_{r}$ (cf. (3.2)).  To see that (3.6) holds, by (3.7) it suffices to show that 
\begin{equation}
\big(\{ z \, |\, z \in  {\tilde{\cal H}}_{-r}, \, \varphi(z) <t \} \cap {\tilde{\cal Y}}_0 \big)  \in \sigma [{\tilde{\cal Y}}_0], 
\quad {\mbox{for any \, $\varphi \in {\tilde{\cal H}}_{r}$ \, and \, $\forall t \in {\mathbb R}$}}.
\end{equation}
By Lemma 2.1,  since $ {\tilde{\cal H}}_{-r} \supset {\tilde{\cal Y}}_0$, and it holds that
\begin{equation*}
\big( \{ z \, | \, z \in {\tilde{\cal H}}_{-r}, \, \varphi (z) <t \} \cap {\tilde{\cal Y}}_0 \big) 
= \{ z \, | \, z \in{\tilde{\cal H}}_{-r} \cap {\tilde{\cal Y}}_0, \, \varphi (z) < t \} 
= \{ z \, | \, z \in {\tilde{\cal Y}}_0, \, \varphi (z) < t \},
\end{equation*}
we see that (3.8) is equivalent to the following:
\begin{equation}
\{ z \, | \, z \in {\tilde{\cal Y}}_0, \, \varphi (z) < t \} \in \sigma [{\tilde{\cal Y}}_0], 
\quad {\mbox{for any \, $\varphi \in {\tilde{\cal H}}_{r}$ \, and \, $\forall t \in {\mathbb R}$}}.
\end{equation}

On the other hand, by (2.38), from the Sobolev's embedding theorem (cf., e.g., Th. 3.15 in [Mizohata 73]), 
since
$C_b({\mathbb R}^d \to {\mathbb R})  \subset W^{r(d +1),2}({\mathbb R}^d)$,
it holds that 
\begin{equation}
{\tilde{C}} \equiv \{ (|x|^2 + 1)^{- r(d +1)} \psi \, | \, \psi  \in C_b({\mathbb R}^d \to {\mathbb R})   \} 
\supset {\tilde{\cal H}}_{r},
\end{equation}
where $C_b$ denotes the space of real valued bounded continuous functions, and 
$W^{r(d +1),2}$ denotes the Sobolev space defined., e.g.,  by Def. 2.9 of [Mizohata 73],
where the notation such that ${\cal E}_{L^2}^{r(d +1)} = W^{r(d +1),2}$ is adopted.
Thus, by (3.10),  in order to prove (3.9), that is equivalent to (3.8), it suffices to show that 
\begin{equation}
 \{ z \, | \, z \in {\tilde{\cal Y}}_0, \, \varphi (z) < t \} \, \in \, \sigma[{\tilde{\cal Y}}_0], 
\qquad  \quad \forall \varphi \in {\tilde{C}}, \quad \forall t \in {\mathbb R}.
\end{equation}
For (3.11), we used the fact that ${\tilde{C}}$ can be taken as the dual space of ${\tilde{\cal Y}}_0$, which is included in the proof of Lemma 2.1 (cf. (3.2)), but is easily seen as follows:
 By (3.10), for $\varphi = (|x|^2 + 1)^{- r(d +1)} \psi \in {\tilde{C}}$ with $\psi \in C_b$ and any $z \in {\tilde{Y}}_0$,   it holds that 
\begin{eqnarray}
  \lefteqn{
      |\varphi(z)| = |\sum_{i =1}^{\infty} m_i \varphi(y_i) | = | \sum_{l =0}^{\infty} \big( \sum_{y_i \in Q_t: |t| =l} m_i \varphi(y_i) \big)| }   \nonumber \\
     &&\leq \sum_{l=0}^{\infty} \big( \sum_{|t| = l} n({\mathbb Y}, t) (\sup_{y \in  Q_t} |{\varphi}(y)|) \big) 
  %\leq \sum_{l=0}^{\infty} \big( \sum_{|r| = l} (n^2({\mathbb Y}, r)) (\sup_{y \in  Q_r} |{\varphi}(y)|) \big)  \nonumber \\
\leq   \| \psi \|_{L^{\infty}} \sum_{l=0}^{\infty} \big( \sum_{|t| = l} (n^2({\mathbb Y}, t)) (\sup_{y \in  Q_t} 
| (|y|^2 + 1)^{-r(d + 1)} \big) \nonumber \\
&&\leq 
\big\{
\| \psi \|_{L^{\infty}} \sum_{l=1}^{\infty} \big( \sum_{|t| = l} (n^2({\mathbb Y}, t)) ((t^2-\frac12)^{-r(d+1)} \big)  + n^2({\mathbb Y},0) \big\} < \infty,
     \end{eqnarray}
where the last equality follows from (2.24) and (2.27).

In addition, note that for $\varphi \in {\tilde{C}}$, by the decomposition such that 
$\varphi = \varphi_+ - \varphi_{-}$, 
 where $\varphi_+(x) \equiv \max \{ \varphi (x), \, 0 \}$ and $\varphi_{-} (x) \equiv \max \{ -\varphi (x), \, 0 \}$,
it holds that 
\begin{equation}
\varphi_+, \, \, \varphi_{-} \, \in {\tilde{C}}.
\end{equation}
Also, note that the following holds:
\begin{eqnarray}
\lefteqn{
\{ z \, | \, z \in {\tilde{\cal Y}}_0, \, \varphi (z) < t\} 
    } \nonumber \\
      &&= \{ z \, | \, z \in {\tilde{\cal Y}}_0, \, 
          \varphi_+(z) - \varphi_{-}(z) < t \}        \nonumber \\
       &&= \bigcup_{s \in {\mathbb Q}} \big( \{ f \, |\, z \in {\tilde{\cal Y}}_0, \, \varphi_+(z) < t +s\} \cap \{z \, | \, z \in {\tilde{\cal Y}}_0, \, \varphi_{-}(z) > s \} \big),
\end{eqnarray}
where ${\mathbb Q}$ denotes the field of rational numbers.
Thus, since the right hand side of (3.14) is a countable operation, from (3.13) and (3.14), to prove (3.11) it suffices to show that the following holds:
\begin{equation}
\{ z \, | \, z \in {\tilde{\cal Y}}_0, \, \varphi(z) < t \} \in \sigma[{\tilde{\cal Y}}_0], 
\quad {\mbox{for any $\varphi \in {\tilde{C}}$ such that $\varphi(x) \geq 0$, $x \in {\mathbb R}^d$, and $\forall t \in {\mathbb R}$}}.
\end{equation}

To this end for $\varphi \in{\tilde{C}}$ with $\varphi (x) \geq 0$, $\forall x \in{\mathbb R}^d$, define $\varphi_n \in {\tilde{C}}$, $n \in {\mathbb N}$, that satisfy the following:
\begin{equation}
0 \leq \varphi_n(x) \leq \varphi_{n +1}(x) \leq \varphi(x), \qquad \forall x \in {\mathbb R}^d, \quad \forall n \in {\mathbb N},
\end{equation}
\begin{equation}
{\rm{supp}}\,[{\varphi_n}] \, \subset \, \{ x \, | \, x \in {\mathbb R}^d, \, |x| \leq n \},
\end{equation}
\begin{equation}
\lim_{n \to \infty} \| \varphi_n - \varphi\|_{L^{\infty}} = 0,
\end{equation}
then, since $z \in {\tilde{\cal Y}}_0$ is a {\it {non-negative (integer)}}-valued Radon measure 
on ${\mathbb R}^d$ (cf. (2.30)), we can use an argument of a  {\it{monotonicity}},   we have
\begin{equation}
\{z \, | \, z \in {\tilde{\cal Y}}_0, \, \varphi(z) < t \} \, = \, \bigcup_{n \in {\mathbb N}} 
\{ z \, | \, z  \in {\tilde{\cal Y}}_0, \, \varphi_n (z) < t \}.
\end{equation}
For each ${\varphi}_n \in C_0({\mathbb R}^d \to {\mathbb R}_+)$, there exists a sequence of {\it{simple functions}} $\{{\varphi}_{n,k} \}_{k \in {\mathbb N}}$ on 
${\cal B}({\mathbb R}^d)$, the Borel $\sigma$-field of ${\mathbb R}^d$, such that 
\begin{equation}
0 \leq {\varphi}_{n,k} (x) \leq {\varphi}_{n, k+1} (x) \leq {\varphi}_n(x), \qquad 
\forall x \in {\mathbb R}^d, \quad k \in {\mathbb N}, 
\end{equation}
\begin{equation}
\lim_{k \to \infty} \|  {\varphi}_{n,k}  - {\varphi}_n \|_{L^{\infty}} = 0,
\end{equation}
where $C_0({\mathbb R}^d \to {\mathbb R}_+)$ denotes the space of non-negative continuous functions on ${\mathbb R}^d$ with compact supports.
Then, by (3.20), (3.21) and again by the monotonicity of the sequence of sets, that follows from the {\it{positivity}} of $z \in {\tilde{\cal Y}}_0$, it holds that
\begin{equation}
\{z \, | \, z \in {\tilde{\cal Y}}_0, \, {\varphi}_n(z) < t \} \, = \, \bigcup_{k \in {\mathbb N}} 
\{z \, | \, z \in {\tilde{\cal Y}}_0, \, {\varphi}_{n,k}(z) < t \}.
\end{equation}
 By the definition of the $\sigma$-field $\sigma[{\tilde{\cal Y}}_0]$, provided through 
(2.21), (2.28) and (2.32), since
\begin{equation*}
\{z \, | \, z \in {\tilde{\cal Y}}_0, \, {\varphi}_{n,k}(z) < t \} \, \in \, \sigma[{\tilde{\cal Y}}_0], 
\qquad \forall n \in {\mathbb N}, \quad \forall k \in {\mathbb N},
\end{equation*}
from (3.22) we have 
\begin{equation*}
\{z \, | \, z \in {\tilde{\cal Y}}_0, \, {\varphi}_{n}(z) < t \} \, \in \, \sigma[{\tilde{\cal Y}}_0], 
\qquad \forall n \in {\mathbb N},
\end{equation*}
and thus, from (3.19), 
 we see that (3.15) holds.  This complete the proof of (3.6).
 \qquad
\bsquare

\medskip

{\small{\bf{Acknowledgements}}  
The authors would like to gratefully acknowledge 
the support received from 
 various institutions and grants.  In particular 
for the first and the fifth named author, the conference "Random transformations and invariance in stochastic dynamics" held in Verona 2019 supported by 
Dipartimento di  Matematica, Universit{\`a} degli Studi di Milano, and its organizers, in particular  prof. S. Ugolini, where they could get fruitful discussions  with the participants, e.g., prof. F. Guerra, prof. P. Blanchard, prof. D. Elworthy to whom strong acknowledgements are expressed; 
for the fourth and fifth named authors, IAM and HCM at the University of Bonn, Germany; also for the fourth and fifth named authors, 
international conference
"mathematical analysis and its application to mathematical physics" held at
Samarkand Univ., 2018 supported by Samarkand Univ. Uzbekistan, and its organizer  prof.
S. N. Lakaev ;
for the fifth named author, SFB 1283 and Bielefeld University, Germany; also for the fifth named author, 
the conference "Quantum Bio-Informatics" held at Tokyo University of Science 2019 supported by Tokyo University of Science, and its organizer prof. N. Watanabe, where he could get fruitful discussion with prof. L. Accardi to whom a strong acknowledgement is expressed; moreover for the fifth named author, 
the conference " Stochastic analysis and its applications" held at Tohoku Univ. supported by 
the grant 16H03938 of Japan Society for the Promotion of Science, and its  organizer prof. S. Aida.
Also, the fifth named author expresses his strong acknowledgements to prof. Michael R{\"o}ckner 
for several fruitful discussions on the corresponding researches.
}

                                                                %\begin{figure}
                                                                %\includegraphics{example.eps}
                                                                %\end{figure}
                                                                %\begin{figure*}
                                                                %\includegraphics[width=0.75\textwidth]{example.eps}
                                                                %\end{figure*}
                                                  %\begin{table}
                                                  %\begin{tabular}{lll}
                                                  %\hline\noalign{\smallskip}
                                                  %\end{tabular}
                                                  %\end{table}

   %{\small{\bf{acknowledgements}}   }


%{\bf{References should be extended and corrected.}}

\end{document}